\documentclass[11pt,a4paper]{article}
\usepackage[english]{babel}
\usepackage{amssymb,amsmath,amsbsy}
\usepackage{verbatim}
\usepackage[latin1]{inputenc}
\usepackage[pdftex,bookmarks]{hyperref}
\usepackage{pdflscape}
\usepackage{anysize}
\usepackage[all]{xy}
\marginsize{3cm}{3cm}{3cm}{3cm}
\title{\huge{A Bond Consistent Derivative Fair Value}}

\author{C. Johan Gunnesson\thanks{cje.gunnesson@bbva.com}\qquad Alberto Fern\'andez Mu\~noz de Morales\thanks{a.fernandez.munozmor@bbva.com} \\BBVA - Risk Methodologies\thanks{The opinions of this article are those of the authors and do not reflect in any way the views or business of their employer.}}

\begin{document}

\maketitle

\begin{abstract}

In this paper we present a rigorously motivated pricing equation for derivatives, including cash collateralization schemes, which is consistent with quoted market bond prices. Traditionally, there have been differences in how instruments with similar cash flow structures have been priced if their definition falls under that of a financial derivative versus if they correspond to bonds, leading to possibilities such as funding through derivatives transactions. Furthermore, the problem has not been solved with the recent introduction of Funding Valuation Adjustments in derivatives pricing, and in some cases has even been made worse.

In contrast, our proposed equation is not only consistent with fixed income assets and liabilities, but is also symmetric, implying a well-defined exit price, independent of the entity performing the valuation. Also, we provide some practical proxies, such as first-order approximations or basing calculations of CVA and DVA on bond curves, rather than Credit Default Swaps.

\end{abstract}

\newpage

\setlength{\baselineskip}{1\baselineskip}

\section{Introduction and Final Pricing Formula}

Ever since Black, Scholes and Merton's seminal works \cite{BlackScholes,Merton}, and until recently, financial derivatives products have been priced without taking into consideration credit- or funding spreads of either counterparty in the transaction. A frequently stated reason for this approach (eg, \cite{HullWhite2013a}) was that financial institutions could, in the pre-financial crisis world, borrow funds at the prevailing Libor rate, and any funding considerations could therefore be taken into account by discounting cash flows accordingly. 

This did, however, not correctly reflect the counterparty credit risk inherent in any given derivative. A digital option which is far in the money behaves similarly to a zero-coupon bond, yet traditionally its cash flows were discounted at Libor, or a similar rate, instead of applying the corresponding bond curve. The inconsistency in the way these two, functionally similar, deals were treated led some market participants to securing funding through derivatives transactions. Derivatives desks are of course aware of current bond prices, and it is therefore expected that they should charge the counterparties accordingly, reflecting the value of the liquidity provided to the counterparty. But the accounting mismatch still provided incentives for closing deals entailing funding, thus rendering an upfront profit for the dealer, while at the same time reducing funding costs for the counterparty. It is true that, from the risk management side, some sophisticated banks were provisioning expected counterparty credit losses based on market-implied estimates. However, many were basing such provisions on historical data. 

A step in the direction of reconciling bond- and derivative valuations was taken with the entry into force of the accounting standard IFRS 13. This standard defines fair value as an exit price, further stressing the use of market-implied (or at least market-adjusted) valuations, and including a bank's own non-performance risk, ie, the possibility that the bank may not fulfill all of its obligations. The fair value should not be entity-specific, in the sense that other market participants should arrive at the same valuation. The standard interpretation of IFRS 13 is to include in the derivative price a Credit Valuation Adjustment (CVA), representing the market value of the deal's counterparty credit risk\footnote{The market value of a given risk can be defined as the cost of buying protection against it in the market, ie, the cost of hedging it. It is true that CVA is often expressed as the expected value of discounted losses due to counterparty defaults, but it should be born in mind that these expectations are based on market inputs, and not on historical (or real-world) losses due to counterparty defaults.}, together with a Debit Valuation Adjustment (DVA), representing the bank's non-performance risk and based on its credit spread. The fair value obtained in this way is symmetric, in the sense that two counterparties will arrive at the same value if they use the same calculation methodology and market inputs. For a technical account of these subjects see \cite{Gregory2009,Brigo2009}, or the textbooks \cite{Gregory2012,KenyonStamm2012,Brigo2013}.

After including CVA and DVA, deals with risky counterparties that once seemed artificially appealing will not produce as large an accounting profit upfront, thus reflecting the true nature of these transactions. The most frequent way to quantify CVA and DVA is to estimate market-implied default probabilities using Credit Default Swaps (CDS), instruments in which an insurance premium, the CDS spread, is exchanged for protection against losses stemming from a given bond issuer's default. With equal recovery rates, a higher CDS spread entails a greater probability of default. Credit risk is not the entire story however. A persistent property of bond markets is the existence of a difference between the excess rates of return of bonds over the risk-free rate\footnote{It is, of course, doubtful that any truly risk-free interest rate can be said to exist, but for practical (and theoretical, as will be discussed below) purposes an Overnight Indexed Swap (OIS) rate is often employed (see \cite{HullWhite2013a}). It has become a standard to pay such rates for held collateral, and they are therefore often referred to as collateral rates.} plus the CDS spread. Nevertheless, a bond and a CDS on the same reference basically refer to the same type of risk, so in a frictionless market, arguments of arbitrage should end up driving such difference, called the bond-CDS basis, to zero. There are a number of reasons that explain why this gap fails to disappear completely, besides the classical argument of capital constraints preventing arbitrage opportunities (see \cite{ShleiferVishny}). Certain frictions, like the Cheapest-to-deliver option embedded in CDS contracts (\cite{Blanco2005,Jankowitsch}) or haircuts that the arbitrageur encounters when financing bond purchases in the repo market (\cite{Garleanu}), explain the rationale behind the basis. For further details on the origin of the basis see, for example, \cite{Augustin}. For simplicity, the many reasons underlying the basis are commonly referred to as liquidity risk.

In the past couple of years, an additional step has been taken by some sophisticated banks, with the inclusion of a Funding Valuation Adjustment\footnote{See, for example, the aforementioned textbooks or \cite{MoriniPrampolini2010,PallavaciniBrigo2011,PallavaciniBrigo2012}.} (FVA). The aim of such an adjustment is to take into consideration the funding costs associated to the ''production'' of a derivative's transaction, ie, the cost of funding the hedging of its risks during the lifetime of the deal. This introduces the bank's complete bond spread into the derivatives price. However, rather than solving the discrepancy between bond- and derivatives prices, many approaches to FVA actually make it larger. Consider, for example, the in-the-money digital option mentioned above, and suppose that the bank has bought the option, analogously to a bond purchase. Any approach to pricing it consistently with bonds should thus contain a CVA, reflecting the counterparty's credit risk, together with an additional term governed by the counterparty's bond-CDS basis to reflect the bond's liquidity premium. In contrast, one approach to FVA adds the bank's complete funding cost (proportional to its funding spread) to the calculated CVA. The counterparty's bond-CDS basis is therefore not part of the price and furthermore, as we will explain later, its CVA contributes implicitly to the bank's funding spread, and is therefore double counted. Needless to say, the obtained valuation will not be symmetric, and the counterparty will calculate a different derivatives price. In Section \ref{sec:BondDerivativeDivide} we will explain the limitations of current FVA frameworks in more detail.

In this paper we provide a solution in the form of a rigorously motivated derivative pricing equation that is completely consistent with market bond prices. We will be concerned here with uncollateralized derivatives, since the direct exposure that they generate to the counterparty is analogous to bond exposure, although we will briefly comment on the partially collateralized case in Section \ref{sec:PartiallyCollateralized}. At a given time $t$, the pricing equation takes the form
\begin{equation}
V_t = V_t^c - CVA_t + DVA_t +BFVA_t \ ,
\label{eq:FinalPricingEquationForm}
\end{equation}
where $V_t$ is the fair value at $t$, and its components are
\begin{itemize}

\item $\boldsymbol{V_t^c}$ : the fair value that would be obtained at $t$ if the derivative were perfectly collateralized, meaning that collateral is posted in a continuous fashion by the bank or counterparty in response to changes in the derivative valuation. In \cite{Piterbarg} it was shown that in such idealized cases the derivative value is simply obtained by discounting all future cash flows using the rate paid on the collateral accounts.

\item $\boldsymbol{CVA_t}$: The CVA calculated to adjust for the counterparty's credit risk. 

\item $\boldsymbol{DVA_t}$: The DVA reflecting own credit risk, and which equals the CVA that would be calculated by the counterparty.

\item $\boldsymbol{BFVA_t}$: The new term in our approach, which we call a \textbf{Bilateral Funding Valuation Adjustment}, and that incorporates the effects of both the bank's and counterparty's bond-CDS bases. In turn, we separate it as
\begin{equation}
BFVA_t = -CFVA_t + DFVA_t \ , 
\end{equation}
where $\boldsymbol{CFVA_t}$ stands for \textbf{Credit Funding Valuation Adjustment}, and is governed by the counterparty's bond-CDS basis, while $\boldsymbol{DFVA_t}$ means \textbf{Debit Funding Valuation Adjustment}, and depends on the bank's bond-CDS basis. They can be thought of as correcting CVA and DVA, respectively, extending them to a full funding adjustment. In particular, positive exposure to the counterparty, given by
\begin{equation}
V^+_t \equiv \max (V_t,\, 0) ,
\end{equation}
and which arises when the derivative can be considered an asset, generates $CFVA_t$, while negative exposure (the derivative is a liability)
\begin{equation}
V^-_t \equiv \max (-V_t,\, 0) 
\end{equation}
gives rise to a $DFVA_t$. In more detail,
\begin{equation}
CFVA_t = E\left[ \int_t^T 1_{\text{alive}}(s)\, D(t,\,s)\, \gamma^C_s \, V^+_s \, ds \right] \ ,
\label{eq:CFVASimple}
\end{equation}
where $E[\cdot]$ stands for ''Expected Value\footnote{Under the risk-neutral, or market-implied, measure.}'', $t$ is the current valuation time, $T$ is maturity of the deal, $s$ is an integration variable representing all intermediate times between the present ($t$) and maturity ($T$), $1_{\text{alive}}(s)$ means that the deal should be alive at $s$ (it is a variable that is equal to 1 if the deal is alive at $s$, and zero otherwise), $D(t,\,s)$ is the discount factor between $t$ and $s$, $\gamma^C_s$ is the counterparty's bond-CDS basis at $s$, and $V^+_s$ is the aforementioned positive exposure. In the same way,
\begin{equation}
DFVA_t = E\left[ \int_t^T 1_{\text{alive}}(s)\, D(t,\,s)\, \gamma^B_s \, V^-_s \, ds \right] \ ,
\label{eq:DFVASimple}
\end{equation}
where $\gamma^B_s$ is the bank's bond-CDS basis.

\end{itemize} 

It should be noted that the pricing equation \eqref{eq:FinalPricingEquationForm} is symmetric, and as a consequence price agreement between two counterparties using it will always be possible. The proposal is therefore especially suitable in an accounting framework, since price agreement implies that it should always be possible to exit the deal at that price\footnote{Recently, a similar result has been obtained in \cite{Lou}, starting from rather different assumptions. In this paper, the author posits that counterparties, rather than remaining in an uncollateralized setup, have economic incentives to willingly enter into collateral agreements in which the rate paid on received collateral is their funding rate. The different rate paid on collateral then leads to funding adjustments similar to our \eqref{eq:CFVASimple} and \eqref{eq:DFVASimple}.}.

In the case that the derivative is partially collateralized, the resulting expressions remain relatively simple, assuming cash-collateralization. If we name $C(t)$ the amount of collateral held by the bank at time $t$ (which is defined as negative if the bank has posted net collateral), we get

\begin{equation}
CFVA_t = E\left[ \int_t^T 1_{\text{alive}}(s)\, D(t,\,s)\, \gamma^C_s \, (V_s - C(s))^{+} \, ds \right] \ ,
\label{eq:CFVASimpleColl}
\end{equation}
and
\begin{equation}
DFVA_t = E\left[ \int_t^T 1_{\text{alive}}(s)\, D(t,\,s)\, \gamma^B_s \, (V_s - C(s))^{-} \, ds \right] \ .
\label{eq:DFVASimpleColl}
\end{equation}

In the next section we will bring into focus the limitations of current approaches to derivative pricing and FVA. We will then detail and motivate our proposal for a derivative fair value in Section \ref{sec:ReconcilingTwoWorlds} and discuss briefly the (partially) collateralized case in Section \ref{sec:PartiallyCollateralized}. We finish with conclusions and a final discussion in Section \ref{sec:Conclusions}. In the appendices, a mathematically rigorous derivation of the pricing formula is provided in Appendix \ref{app:DerivationOfPricingEquation}, while an analytical expression for bond prices, allowing a calibration to market prices, can be found in \ref{app:BondConsistency}.

\section{The Bond-Derivative Divide}

\label{sec:BondDerivativeDivide}

Before continuing in Section \ref{sec:ReconcilingTwoWorlds} with our proposal, we will now highlight some of the current approaches to derivative pricing and FVA, and in particular focus on the discrepancy between bond- and derivative valuations. This section is independent of the rest of the paper, but should shed some light on the validity of the pricing equation \eqref{eq:FinalPricingEquationForm}.

The alternative approaches that we will discuss are $FVA=0$, own bond-CDS based FVA, transfer cost-based FVA, CVA and full FVA with no DVA, asymmetric FVA with only a funding cost, and CVA calculated using bond curves. In many cases, the inclusion of FVA will make the discrepancy with bond prices worse, entailing for example double counting of Credit- and Funding Valuation Adjustments.

\subsection{Should FVA be zero?}

In \cite{Hull2012a} the controversial claim was made that derivative pricing should include CVA and DVA, but not FVA. In \cite{Hull2012b} and \cite{Hull2013b} the same authors admit that it is only defensible to base an FVA on the bank's bond-CDS basis, but they also argue that it does not seem compatible with IFRS 13 to include such an FVA in the accounting fair-value, since it would consist of an entity-specific valuation. Arguably, different counterparties will arrive at different prices.

Others have recognized that it seems to make economic sense to calculate even a substantial FVA, reflecting the costs of funding the derivative's hedges, and take it into consideration in decision-making processes, but that it should not be used for accounting purposes \cite{Ruiz2013}. The reason given is that the market price is not fixed by a given entity's funding costs, in the same way that the market price of a given commodity is not a direct function of a given producer's extraction costs. Yet another interpretation of this observation is that a ''market'' FVA should be calculated, based on the average funding costs of different market actors.

However, it is not correct that IFRS 13 implies that valuations cannot be based on parameters depending on the own entity, as the inclusion of DVA exemplifies. Instead, the parameters should be such that agents other than the entity will arrive at the same values. In other words, they should be based on objective, market-based information. Consistency with bond prices requires the introduction of an FVA, but as long as it is calculated using market information, there is no problem in including it in an accounting fair value.

In \cite{BurgardKjaer2011,BurgardKjaer2012} a zero FVA is also obtained under the assumption that the bank can freely trade in own bonds of different seniority. In fact, the methodology that we use in this paper, that of replication, follows the same principles as that of these papers. The result of this pricing methodology depends on the risk factors on which the product value is assumed to depend, and if a dependence on the full funding spread is required, as it must if consistency with bond prices is required, the bond-CDS basis will emerge from the replication.

\subsection{Incorporating own bond-CDS basis}

In the previous case we mentioned that \cite{Hull2012b} and \cite{Hull2013b} allowed for an FVA depending on the bank's bond-CDS basis, although the authors expressed doubts whether such an FVA is compatible with accounting standards. Such a dependence was first proposed in \cite{MoriniPrampolini2010}, and in \cite{GunFer2014} we arrived at a similar result by different means, although we do not believe that the accounting mismatch is irreconcilable. The only problem with these approaches, in our view, is that they do not incorporate the counterparty's funding spread, and can therefore not reproduce the market prices for the counterparty's bonds.

\subsection{Transfer cost-based FVA}

From the perspective of managing a derivative desk it seems appealing to base FVA on the bank's internal funds transfer rates, since they determine the actual financing costs experienced by the desk. In Proposition 3 of the report \cite{KPMG2013}, it is stated that it is desirable to have a close alignment between FVA calculations and funds transfer pricing rules, in order to avoid arbitrage opportunities for external liquidity takers. A client should not be able to benefit from entering an in-the-money derivative as compared to taking a traditional loan, it is argued.

From the accounting perspective it makes less sense to base FVA on funds transfer rates. Clearly, doing so would constitute an entity-specific calculation, and not be directly market based. And the problem goes deeper as banks that can freely alter their transfer rates could directly manipulate their obtained accounting FVA. It should instead be the transfer costs that should adapt to the external conditions, with FVA calculated independently. Finally, if the goal is to avoid arbitrage opportunities for external liquidity takers it becomes even more relevant to impose consistency between funding through derivatives and bonds.

\subsection{Unilateral CVA and full FVA with no DVA}
\label{sec:CVAplusFullFVA}

As summarized in Proposition 8 of the report \cite{KPMG2013}, DVA is difficult to monetize, impossible to completely hedge, and from the point of view of the derivatives desk leads to a negative carry ie, a loss of value with the passage of time. This has led to proposals excluding DVA from pricing, and instead including a full FVA, that is, an FVA based on the bank's full funding spread. In Section 14.3.6 of \cite{Gregory2012}, a first-order approach in this spirit is presented, denoted as CVA+FCA+FBA. Furthermore, in \cite{Garcia2013} such a price was obtained via replication, including higher order corrections, by not considering the bank's default as a risk-factor.

In the case of liabilities this approach will yield a numerically similar result as our proposal \eqref{eq:FinalPricingEquationForm}, since an FVA based on the complete funding spread is at first order equal to DVA plus an FVA based on the bond-CDS basis, but there will be large differences for assets. The reason for this is that in the CVA + full FVA setup there is an important double counting of CVA. Following the example given in Proposition 6 of \cite{KPMG2013}, if the bank were to deal with a single counterparty, the bank's riskiness, and therefore its funding spread, should be entirely determined by the riskiness of the counterparty. Implicit in the bank's funding spread is therefore the counterparty's CVA and adding the two will produce a full double counting. Proponents of this approach argue that both the CVA hedge and the liquidity provided to the counterparty must be financed independently. However, if the bank were to hedge its exposure to the counterparty, what is left is a purely riskless instrument, and the bank's funding spread should adjust accordingly, since what remains will be a riskless balance sheet. In other words, hedging CVA does not have to be considered a cost, but can instead be thought of as an investment in risk reduction, accompanied by the associated funding benefit. 

It is true that funding costs will not adapt instantly to variations in risk profiles. However, as discussed in \cite{GunFer2014}, large portions of the bank's derivatives portfolio are continuously renewing, with similar deals replacing maturing ones. For such a steady state balance sheet, the funding spread should have adapted to the riskiness of the portfolio. In the case of an expanding or contracting business the situation is more complex, but there will in general still be a significant overlap between the portfolio compositions in, for example, consecutive months, implying that funding costs should reflect the assumed risks.

The use of CVA + full FVA means that credit given to the counterparty through derivatives transactions will be charged a rate greater than the market rate. Of course, an institution should always try to maximize profits, and if a counterparty is prepared to accept such a price the bank should agree, but we would argue that such a scenario should lead to an upfront accounting gain. The risk of using CVA + full FVA is instead that deals that are actually profitable may not appear as such, leading to lost business opportunities. Furthermore, the negative carry of DVA is actually not real\footnote{This is at least the case for a steady state balance sheet.} since DVA can be identified with a funding benefit. If this benefit is compensated for internally, the negative carry will go away.

\subsection{Asymmetric FVA with no funding benefit}
 
Yet another possibility, presented in Section 14.3.6 of \cite{Gregory2012}, is to calculate a CVA + DVA + FCA, where FCA (the funding cost) is the part of a full FVA that is based on positive exposures. This has the advantage, from the management point of view, that FCA, being based on the bank's complete funding spread, constitutes a natural hedge for the DVA term. However, numerically the approach is similar to CVA + full FVA, and therefore has the same problems. There will be a mismatch between credit provided to the counterparty through derivatives versus by other means.

\subsection{Calculating CVA and DVA from bond curves}
\label{sec:CVADVAFromBondCurves}
 
In this section we have taken a critical position regarding common FVA frameworks, but let us end it on a positive note. Some financial entities have reported calculating a DVA term based upon the default probabilities extracted from their own issued bonds. By doing so, they argue, they do not need to calculate any FVA term, since such a DVA also encompasses their funding capabilities. Ordinarily, CVA and DVA depend on the credit component of the counterparty's and bank's respective funding spreads, and the question is therefore whether one can incorporate funding considerations by extending them to the full funding spreads. In other words, we would infer the default probabilities underlying these adjustments from bonds rather than CDSs. 

In fact, such an approach is not unreasonable, and if it is applied to both CVA and DVA, as we show in Appendix \ref{app:AFirstOrder}, it can be obtained as an approximation to the exact pricing equation obtained in the next section. Furthermore, its non-recursive form makes it much simpler to implement, making it a practical proxy for the full calculation.
 
\section{Reconciling the two worlds: Replicating Derivatives with Bonds}

\label{sec:ReconcilingTwoWorlds} 

In this section we will motivate our proposal. For a more rigorous derivation see Appendix \ref{app:DerivationOfPricingEquation}. We will start with the so-called perfectly collateralized case in Section \ref{sec:PerfectlyCollateralized}, for which there is neither CVA, DVA nor FVA, and we will then explain how the general case arises by modeling the financing of the absence of collateral in terms of bonds.

\subsection{A perfectly collateralized point of departure}

\label{sec:PerfectlyCollateralized}

The point of departure for the general case will be a perfectly collateralized trade, a theoretical construct in which there is a continuous exchange of cash collateral so that at all times neither counterparty has any net exposure to the other. This implies that there will be no losses due to defaults, and therefore no CVA or DVA. What is perhaps less obvious is that there will not be any funding adjustments either. The reason is that it is assumed that the trade can be hedged in the interbank market using (different) collateralized trades. The initial cost of setting up hedges and residual balances in cash accounts should coincide with the premium paid by the counterparty. Furthermore, any variation of the MtM of the original deal implying, for example, that the bank must post additional collateral to the counterparty will be offset by collateral posted to the bank by the counterparties of the hedges.

As shown in \cite{Piterbarg}, the value of such a perfectly collateralized derivative is obtained by discounting all future cash flows by the rate paid on the held collateral, which in typical collateral agreements is taken to be an OIS rate $c$ based on overnight interbank lending. To understand why this discount rate should be used, consider a deal in which the bank receives a single fixed cash flow $N$. An instant before maturity, the deal value will be equal to $N$, since any discount can be ignored, and the cash held in collateral accounts will also equal $N$. At maturity, the deal is closed, but there is no net exchange of cash between the bank and the counterparty since the final cash flow $N$ will cancel the return of the collateral. Now, at previous times, the bank must pay the OIS rate $c$ on the collateral account balance, but since there will be no net cash exchange at maturity it will only agree to paying this rate if at the same time the collateral balance increases by the same amount, compensating this outflow (so that there are no net cash interchanges at previous times either). 

The end result is that the amount of collateral, and therefore the value of the deal, grows at the rate $c$. In other words, previous values of the deal are obtained by discounting using precisely $c$. If we denote the deal value at time $t$ by $V_t^c$, maturity by $T$, and taking into account the continuous accruing of the collateral account, we therefore have 
$$V^c_t = \exp^{-(T-t)\cdot c}\, N \ .$$ 
Allowing for the possibility that $c$ may vary in time (although in a way known beforehand) as $c_t$, this becomes\footnote{Let $t_1=t,\, t_2,\ldots ,\, t_n=T$ be times between which the collateral rate $c$ is fixed, and which define the periods $\Delta t_i \equiv t_{i+1}-t_i$ upon which the collateral account interest rate payments are based (typically daily periods). Ignoring subtleties regarding day-count conventions, the collateral account will then grow between $t_i$ and $t_{i+1}$ by a factor $(1+c_{t_i}\cdot \Delta t_i)$. Using that $\Delta t_i$ is small for frequent collateral margining, this factor can be substituted for $\exp \left(c_{t_i}\cdot \Delta t_i \right)$, since the former expression is the first order approximation of the latter.  Multiplying the factors stemming from all periods gives $V_T^c = \prod_i^{n-1} \exp \left(c_{t_i}\cdot \Delta t_i \right) V_t^c  = \exp \left(\sum_i^{n-1}c_{t_i}\cdot \Delta t_i \right) V_t^c$. Substituting the sum inside the exponential for an integral (exact in the limit when $\Delta t_i \rightarrow 0$) gives the result.  }
$$V^c_t = \exp ^{-\int _t ^T  c_s ds}\, N \ .$$ 

Of course, in reality the values of $c_s$ at times later than $t$ are not known at $t$, so we need to take expectations over the different paths that the collateral rate can take\footnote{The exact measure needed to take this expectation is obtained from a more careful analysis, see \cite{Piterbarg}, or Appendix \ref{app:DerivationOfPricingEquation}.}:
$$V^c_t = E\left[ \exp ^{-\int _t ^T  c_s ds} \right] \, N \ .$$ 
Following, for example, \cite{BrigoMercurio}, we introduce the stochastic discount factor 
\begin{equation}
D(t,\,t')\equiv \exp\Big(-\int _t ^{t'}  c_s ds\Big).
\label{eq:StochasticDiscountFactor}
\end{equation}

In general, since we are dealing with derivatives, we should expect the final cash flow to be an unknown payoff, following some statistical distribution, possibly correlated with the discount factor. This payoff, occurring at maturity, can be written simply as $V^c_T$, and we therefore have
\begin{equation}
V^c_t = E\left[ D(t,\, t') V^c_T \right]  \ .
\end{equation}

\subsection{Financing the collateral gap}

Let us now turn to the case of main interest, in which the derivative is uncollateralized, leaving a discussion of the partially collateralized case for Section \ref{sec:PartiallyCollateralized}.

If the deal is not collateralized there will be two consequences: Firstly, there will be losses upon defaults of either the bank or counterparty, depending on the sign of the value of the trade at that moment. Secondly, the lack of posted collateral has to be financed to compensate for the collateral movements of any market hedges, and also the funding implicit in the trades net value. Funding the trades NPV, applying the bank's funding curve, transfer costs, etc, is what has led to previous proposals on FVA. Our main point is that we recognize that in the case of positive exposure to the counterparty, we are financing the counterparty for the amount that would have been transferred as collateral in the perfectly collateralized case. The corresponding funding adjustment thus depends on the accounting value of financing the counterparty, and not the cost of obtaining bank funds. 

Any derivatives trade can be thought of as composed of its perfectly collateralized counterpart together with a financing of any deficit or excess of collateral. If the exposure is positive $V_t^+ > 0$, the counterparty would have transferred posted collateral corresponding to this exposure in the perfectly collateralized case. In allowing it not to do so, the bank is implicitly lending the counterparty funds, or equivalently, buying the counterparty's bonds, whose value we will write as $B^C_i(t)$, with the label '$i$'  specifying the individual bonds. Let $\omega^C_i$ be the quantity held of bond '$i$'.  The total value of these bonds will be precisely $V_t^+$, so we have the \textbf{Credit Constraint}
\begin{equation}
\sum_i \omega^C_i \, B^C_i(t) = V_t^+ \ ,
\label{eq:CreditConstraint}
\end{equation} 
which holds at all times. The bank can hedge this bond component of the derivative by taking a short position in the counterparty's bonds. 

But why is more than one bond needed in \eqref{eq:CreditConstraint}? Could we not simply include a single bond with the same maturity as the derivative? No, unfortunately, the uncertain and possibly bilateral nature of derivative cash flows complicates matters. Consider, for instance, an interest rate swap with a value of 0 at inception, implying a vanishing positive exposure. Does this, together with \eqref{eq:CreditConstraint}, mean that the bond-component of the derivative is zero as well? Clearly, this cannot be the case, because there is a positive probability that the derivative value in the future will move in favor of the bank, generating a non-zero positive exposure and thereby producing a position in the counterparty's bonds. The possibility of this forward position must be included in the initial valuation, and if the counterparty's bond yield rises, the fair value, from the point of view of the bank, should decrease\footnote{Since no additional cash flows are interchanged as a consequence of the change in bond yield, the bank will be lending funds to the counterparty at a rate cheaper than the new market rate, entailing a loss in value.}. Consider, for example, the case of two bonds, and $V_t^+=0$. The credit constraint then becomes
$$\omega^C_1 \, B^C_1(t)+\omega^C_2 \, B^C_2(t)=0 \ .$$
If the duration of, for instance, the second bond is greater than the first and $\omega^C_2 > 0$ (requiring that $\omega^C_1 < 0$), the total value of the bonds will decrease upon parallel increases of the yield curve. The exact amount of bonds would then be determined by requiring a matching of the bond and derivative sensitivities to the yield curve. Due to the constraint, in order to capture $n$ points of the yield curve, at least $n+1$ bonds must be included.

However, in order to correctly represent the dependence of the derivative on the counterparty's credit risk, besides matching derivative- and bond curve sensitivities, the jump-to-default components ie, the change in value as a consequence of the counterparty's default event, must be captured as well. One way to achieve this would be to include additional bonds in \eqref{eq:CreditConstraint} (in, for example \cite{BurgardKjaer2012} the jump-to-default component was replicated using bonds of different seniority), and simultaneously choose the quantities $\omega^C_i$ so that both bond curve and jump-to-default dependencies take the correct values. Another possibility is to include Credit Default Swaps in the mix, as we do in our derivation in Appendix \ref{app:DerivationOfPricingEquation}, greatly simplifying the calculation. Assuming that such CDS can be entered without a relevant upfront payment, they can be added without changing the amount of credit given to the counterparty, and therefore without affecting \eqref{eq:CreditConstraint}. The choice of whether to replicate the jump-to-default component using additional bonds or CDS should not alter the obtained result.

We repeat the steps outlined above for negative exposure, $V_t^- > 0$. In this case, the bank has transferred less collateral to the counterparty, as compared with the perfectly collateralized case. In analogy with the positive exposure case, the counterparty is financing the bank, and it is therefore as if the bank had issued bonds $B^B_i(t)$, of total value $-V_t^-$ (the value is negative for the bank since they are a liability), which were bought by the counterparty. We therefore have the \textbf{Debit Constraint}
\begin{equation}
\sum_i \omega^B_i \, B^B_i(t) = -V_t^- \ .
\label{eq:DebitConstraint}
\end{equation} 
This component can of course be hedged by trading in the entity's bonds. 

The issue of the bank's jump-to-default is more contentious. The reason is that it is difficult for a bank to replicate its own jump-to-default using real instruments\footnote{For example, in \cite{BurgardKjaer2012} the jump-to-default component is replicated using two bonds of different seniority. However, the authors recognize that there might be a mismatch between the post-default value of the bond portfolio and the derivative due to a lack of control over bond recoveries.}, and hedging it can prove to be pernicious (see for example \cite{Castagna2012}). As a case in point, consider that no bank can sell its own CDS (who would buy protection from a seller on the seller's own default?). However, in \cite{GunFer2014} it was explained that, even in a hedging context, it makes sense to model the existing unhedged jump-to-default component using a fictitious position on a CDS written on the bank, which would\footnote{At least in the case of a steady state balance sheet, in which new deals replace other deals on a continuous basis.} be prepared to pay or receive precisely the market spread for such a CDS, even though it could not enter into it in practice. The reason is that cash-flows occurring at default will affect the bank's recovery, which should alter its funding costs. The additional funding costs or benefits obtained in this way compensate for the CDS premium. Furthermore, from an accounting perspective it is important to fully include own credit spreads, since they should be taken into account in any exit price. The final result is that the bank's own credit risk can be treated symmetrically in the same way as for the counterparty.

In sum, any derivatives deal can be decomposed as
\begin{equation}
\text{Derivative}=\text{Perfectly Collateralized Derivative} + \text{Bonds} + \text{(optionally) CDS} \ .
\end{equation}
The way to proceed is to calculate the amount of bonds and CDS, at current and future times, implied by this decomposition, and derive a pricing equation, as is done in the appendix. Here, we will give a heuristic motivation of the result. 

Firstly, any cash flows occurring at default of either party are precisely those underlying calculations of CVA and DVA. By calculating CVA and DVA we are therefore taking into consideration the deal's credit risk. CVA is contingent, meaning that it is based on counterparty defaults taking place prior to any default by the bank itself. The same happens with DVA. The rationale behind this contingency is that the derivative is liquidated upon the first default, by either counterparty, and so its value will not be affected by defaults beyond the first. The expressions for CVA and DVA are therefore
\begin{equation}
CVA_t = E \left[ 1_{\{\tau^C < \tau ^B \}} D(t,\, \tau ^C) \left( 1 - R^C(\tau ^C) \right) \left(V^c_{\tau ^C}\right)^+ \right]
\label{eq:CVA}
\end{equation}
and
\begin{equation}
DVA_t = E \left[ 1_{\{\tau^B < \tau ^C \}} D(t,\, \tau ^B) \left( 1 - R^B(\tau ^B) \right) \left(V^c_{\tau ^B}\right)^- \right] \ ,
\label{eq:DVA}
\end{equation}
where $\tau ^C$ and $\tau ^B$ are times at which the defaults occur for the counterparty and bank, respectively, $R^C(t)$ and $R^B(t)$ are their recovery fractions, $V^c_{t}$ is the value of the perfectly collateralized equivalent of the derivative\footnote{We adopt the riskless close-out convention, leading to perfectly collateralized exposures being used in the expressions for CVA and DVA. An important point, though, is that the value of future defaults may be incorporated into the close-out claim, put forth upon default, if a so-called risky close-out is performed (see \cite{GregoryGerman,Carver}). However, for unsophisticated counterparties, such as the ones that tend to not have collateral agreements with the bank, we would expect a riskless closeout, in which the claim is based on the riskless value, since such counterparties are not likely to have the ability to calculate CVA and DVA. On the other hand, for sophisticated counterparties it is frequent to have a collateral agreement with daily margin, based on the perfectly collateralized value, so it is natural that a close-out should also be based upon this value.}, $D(t,\,\cdot )$ is the stochastic discount factor \eqref{eq:StochasticDiscountFactor}, and $1_{\cdots}$ is an indicator function requiring the stated condition to be satisfied. Basically, we are taking the expected value of all future losses due to defaults, given by the perfectly collateralized equivalent exposures at the time of default discounted to the present time. The indicators imply that only losses corresponding to the first default will contribute, reflecting the contingency. Usually, we will assume constant recoveries, and so $R^C(t) \equiv R^C$ and $R^B(t) \equiv R^B$.

These terms include the credit risk of the bonds satisfying \eqref{eq:CreditConstraint} and \eqref{eq:DebitConstraint}, so the only additional contribution to the derivative value should depend on the fraction of the bond yield not explained by credit risk. If we decompose a bond yield $f^X_t$, where $X$ is either equal to $C$, for the counterparty, or $B$ for the bank, as
$$
f^X_t = c_t + \pi^X_t + \gamma ^X_t \ ,
$$
where $c_t$ is the OIS rate (often considered as a proxy for a ''risk-free'' rate), $\pi^X_t$ is the CDS spread, encoding credit risk, and $\gamma ^X_t$ the bond-CDS basis (by definition), we are thus saying that the remaining terms contributing to the derivative valuation should be governed by the bond-CDS basis.

In effect, on the credit side we are charging the counterparty for the additional liquidity premium implicit in the funds provided to the counterparty due to the lack of (expected) posted collateral. We will call this term a \textbf{Credit Funding Valuation Adjustment} (CFVA), and not a Funding Cost Adjustment, a common term in the FVA literature, since it should not be confused with the bank's own funding costs. The properties that we expect that this CFVA should have are:
\begin{itemize}

\item Contributions to the adjustment from any future time $s \geq t$ should be proportional to $\gamma^C_s \cdot V^+_s$, since by \eqref{eq:CreditConstraint} the net amount of bonds held are $V^+_s$, and the liquidity premium charged for these bonds is $\gamma^C_s$.

\item Only future scenarios in which the deal is alive should contribute. If either counterparty has defaulted previously, the deal will have been liquidated, and no funding adjustment should be made. This condition can be incorporated by the indicators $1_{\{ \tau^C>s \}}1_{\{ \tau^B>s \}}$.

\item Future contributions to the adjustment should be discounted to the present time using the stochastic discount factor $D(t,\, s)$.

\item We should sum over the contributions to $CFVA_t$ from all future times $s$, implying an integration over the lifetime of the deal.

\item Since the previous discussed factors are stochastic, in general, we should take the expectation of their product (in the appropriate measure) in order to obtain the final adjustment. 

\end{itemize}
In total this gives
\begin{equation}
CFVA_t = E\left[ \int_t^T 1_{\{ \tau^C>s \}}1_{\{ \tau^B>s \}}\, D(t,\,s)\, \gamma^C_s \, V^+_s \, ds \right] \ ,
\label{eq:CFVA}
\end{equation}
and the fair value will be reduced by this amount.

In the same way, we obtain for the debit side the \textbf{Debit Funding Valuation Adjustment} (DFVA), given by
\begin{equation}
DFVA_t = E\left[ \int_t^T 1_{\{ \tau^C>s \}}1_{\{ \tau^B>s \}}\, D(t,\,s)\, \gamma^B_s \, V^-_s \, ds \right] \ ,
\label{eq:DFVA}
\end{equation}
which enters with a positive sign in the fair value.

A nice property of \eqref{eq:CFVA} and \eqref{eq:DFVA} is that they are symmetric: the CFVA calculated by the bank coincides with the DFVA calculated by the counterparty, and vice versa. This prompts us to define a \textbf{Bilateral Funding Valuation Adjustment} BFVA, as
\begin{equation}
BFVA_t = -CFVA_t + DFVA_t \ ,
\label{eq:BFVA2}
\end{equation}
analogously to the Bilateral Credit Valuation Adjustment (introduced in \cite{Brigo2009}). 

Adding up the different contributions, our complete proposal for the fair value $V_t$ of the derivatives transaction is thus
\begin{equation}
V_t = V_t^c - CVA_t + DVA_t +BFVA_t \ ,
\label{eq:FinalPricingEquationForm2}
\end{equation}
with $BFVA_t$ defined by equations \eqref{eq:CFVA}-\eqref{eq:BFVA2}. It should be noted that \eqref{eq:FinalPricingEquationForm2} is recursive, since $V_t$ will depend, through the funding adjustment $BFVA_t$, on $V_s^+$ and $V_s^-$ at later times. This makes the equation difficult and costly to implement for the entire derivatives portfolio, but as we show in Appendix \ref{app:AFirstOrder}, and have discussed in Section \ref{sec:CVADVAFromBondCurves}, reasonable and practical approximation are obtained by either omitting the recursive behaviour of the $FVA$ terms, or by dropping the $BFVA_t$ term altogether and instead incorporating funding considerations into $CVA_t$ and $DVA_t$ by basing them on bond-implied default probabilities.

We have constructed the pricing equation \eqref{eq:FinalPricingEquationForm2} taking into consideration bond-CDS bases in order to be consistent with both the counterparty's and the bank's bonds. But can we actually reproduce observed bond prices in the current framework? Yes, but there are some subtleties when applying so-called recovery conventions. In Appendix \ref{app:BondConsistency} we attempt to shed light on this issue, deriving closed expressions for bond prices under different conventions. The resulting formulae can be used to calibrate the bond-CDS bases to bond market prices.

\section{Partially Collateralized Derivatives}

\label{sec:PartiallyCollateralized}

Even though we will not give a rigorous motivation here, we will dedicate a few lines to explain how the pricing equation can be extended to include partially collateralized derivatives, meaning derivatives subject to a Credit Support Annex (CSA), but cannot be considered perfectly collateralized (for which the value is simply $V_t^c$). Depending on the degree of accuracy required, this may encompass practically all deals, since in reality it is not possible to obtain the ideal of perfect collateralization. We will limit ourselves to cash collateral, noting that non-standard collateral, such as bonds denominated in domestic or foreign currencies, commodities, equities, etc, can be incorporated at the expense of considerable complexity in the pricing (affecting the relevant pricing measures, etc). For an example of the treatment of collateralization in a funding framework, see \cite{PallavaciniBrigo2011,PallavaciniBrigo2012}.

Two ways in which counterparty exposure can arise when cash collateralization is not perfect are
\begin{enumerate}

\item The existence of thresholds, minimum-transfer-amounts, independent amounts and non-continuous remargining will imply that, at any given moment $t$, the amount of held collateral $C(t)$ is not exactly equal to the derivatives value $V_t$.

\item At default of either party, there is typically a period of uncertainty $\Delta$, a so-called cure period (or margin period of risk), in which no further collateral is posted by the defaulting party, but it is not yet legally clear that the default has actually taken place and/or the surviving party has not yet closed out all of the market hedges corresponding to the derivatives trade in question. This incremental lack of collateral at default implies an additional exposure to the defaulting party, and thereby a potentially greater loss.

\end{enumerate}

In order to incorporate partial collateralization the first thing that must be done is to model the collateral function $C(t)$. In general, it can be a rather complex function dependent on the path taken by the derivatives price up to $t$. In many circumstances it is sufficient, however, to use an approximate form. For example, if the collateralization is considered perfect, with the exception of the existence of a bilateral threshold $H$, we have
$$C(t) = \text{sign}(V_t^c)\max \left( |V_t^c|-H ,\, 0 \right) \ ,$$
where the amount of collateral depends on the perfectly collateralized value, as is common in collateral agreements. Another special case, which we have already seen, is of course the uncollateralized case in which $C(t)=0$. Further examples can be obtained in \cite{PykhtinRosen} or \cite{Pykhtin}.

Introducing collateral in the framework is then straightforward. To begin with, it is well-known how to incorporate held collateral into CVA and DVA calculations. For CVA, we have:

\begin{equation}
CVA_t = E \left[ 1_{\{\tau^C < \tau ^B \}} D(t,\, \tau ^C) \left( 1 - R^C(\tau ^C) \right) \left(V^c_{\tau ^C}-C(\tau ^C)\right)^+ \right]
\label{eq:CVACollateral}
\end{equation}
and analogously for DVA. This formula is based on a reduced exposure to the counterparty at the time of default by $C(\tau)$.

Now, any amount of held collateral $C(t)$ obviously reduces the collateral gap that must be financed, so the Credit and Debit Constraints become
\begin{equation}
\sum_i \omega^C_i \, B^C_i(t) = \left(V_t-C(t)\right)^+ \ 
\label{eq:CreditConstraintCollateral}
\end{equation} 
and
\begin{equation}
\sum_i \omega^B_i \, B^B_i(t) = -\left(V_t-C(t)\right)^- \ .
\label{eq:DebitConstraintCollateral}
\end{equation} 
It is therefore not surprising that (and as can be shown performing a more rigorous analysis), the modified Credit- and Debit Funding Valuation Adjustments will be
\begin{equation}
CFVA_t = E\left[ \int_t^T 1_{\{ \tau^C>s \}}1_{\{ \tau^B>s \}}\, D(t,\,s)\, \gamma^C_s \, \left(V_t-C(t)\right)^+ \, ds \right] \ 
\label{eq:CFVACollateral}
\end{equation}
and
\begin{equation}
DFVA_t = E\left[ \int_t^T 1_{\{ \tau^C>s \}}1_{\{ \tau^B>s \}}\, D(t,\,s)\, \gamma^B_s \, \left(V_t-C(t)\right)^- \, ds \right] \ ,
\label{eq:DFVACollateral}
\end{equation}

In these equations it is implicitly contained that the collateral itself can generate exposure to the counterparty. For example, if $C(t)$ is negative, due to some large Independent Amount of collateral, while $V_t \approx 0$, the positive exposure will then be
$$\left(V_t-C(t)\right)^+ \approx |C(t)| \ .$$

The second effect that must be taken into account is that of the cure period $\Delta$. If the period is sufficiently short, in the sense that the potential variation in the derivatives value over the period is small compared to the exposures $\left(V_t-C(t)\right)^\pm$, this effect can be ignored. And in general, we would typically ignore it in the funding terms, since they are given as integrals over time, and the additional funding required during the cure period is therefore relatively small. However, for a sufficiently well-collateralized deal it may become the main contribution to the value of CVA and DVA. In order to incorporate it, we set\footnote{This ignores the discount factor between $\tau^C$ and $\tau^C+\Delta$, which for practical purposes can be neglected.}
\begin{equation}
CVA_t = E \left[ 1_{\{\tau^C < \tau ^B \}} D(t,\, \tau ^C) \left( 1 - R^C(\tau ^C) \right) \left(V^c_{\tau ^C+\Delta}-C(\tau ^C)\right)^+ \right]
\label{eq:CVACollateralCuring}
\end{equation}
with the understanding that if $\tau ^C+\Delta > T$ we redefine $V^c_{\tau ^C+\Delta}$ as $V^c_{T}$. The rationale is simply that we take the default time to be the moment in which collateral is no longer posted, while we are still exposed to the counterparty during the cure period. 

In practice, it can be difficult to solve the partially collateralized pricing equation exactly. Consider for example taking into consideration Minimum Transfer Amounts in the collateral function $C(t)$. This will introduce a path-dependent derivatives price, which is difficult to combine with the recursive nature of the pricing equation (recall that equation \eqref{eq:FinalPricingEquationForm2} is recursive since $V_t$ will depend on future values $V_s$, with $s\geq t$ through the funding terms), which is simplest to treat if the equation can be solved from maturity backwards to the present. 

For the uncollateralized case we mentioned the use of bond curves in the calculation of CVA and DVA as an approximation to the full CFVA and DFVA calculations. Some care must be taken when extending this approach to the partially collateralized case, however. The issue is that the cure period $\Delta$ enters into the CVA term but not CFVA. Calculating a bond-based CVA would thus imply financing an exposure that only appears at default during the entire derivative lifetime. The solution is to split the exposure underlying CVA into two pieces, one which is an exposure during life, and one which is an incremental Exposure At Default (which would contain effects stemming from the cure period), defined as the difference between the exposure entering \eqref{eq:CVACollateralCuring} and that of \eqref{eq:CFVACollateral}. We can then calculate two CVA terms, one based on the exposure during life and the full bond curve, and one based on the incremental EAD and credit spreads. The symmetric treatment should of course be given to DVA.

\section{Conclusions and Discussion}

\label{sec:Conclusions}

The aim of this paper has been to define a fair value for a financial derivatives transaction, entered into by a bank and its counterparty, completely consistent with all available market information, including bond prices. After summarizing the steps taken, we will discuss how the approach is
\begin{itemize}

\item Compatible with the notion of exit price, underlying IFRS 13.

\item Consistent with market bond prices.

\item Not entity specific. 

\item Not based on future costs. We will explain why cost-based derivative valuations are not desirable from an accounting point of view.

\item Free of double-counting between CVA and funding costs.

\end{itemize}

In summary, taking the well-defined case of a perfectly collateralized derivative as a point of departure, we obtain the uncollateralized and partially collateralized cases by modeling the financing of the lack or excess of collateral with respect to this case. For example, if the counterparty should have posted more collateral in the perfectly collateralized case, we recognize that we are lending the missing collateral to the counterparty, which implicitly means that we are holding bonds issued by the counterparty for the same amount. Modeling the bonds in a way consistent with the derivative's dependence on the bond curve, and its behavior in default, provides the pricing equation \eqref{eq:FinalPricingEquationForm2}, containing the perfectly collateralized value, contingent CVA and DVA and a Bilateral Funding Valuation Adjustment. The funding terms consist of a Credit Funding Valuation Adjustment, depending on the counterparty's bond-CDS basis, and a Debit Funding Valuation Adjustment, governed by the bank's bond-CDS basis. Altogether we obtain a symmetric pricing formula implying that both parties will obtain the same valuation, which we have shown to be \textbf{consistent with bond prices}. Due to its recursive nature, the pricing formula may prove difficult to implement, but can be approximated by for instance incorporating the calculation of the FVA terms into CVA and DVA based on bond-implied default probabilities.

From the management point of view it might not be desirable for a deep in-the-money derivative to behave like one of the counterparty's bonds, since the cost of funding of collateral depends on the bank's funding spread. However, the counterparty bond component of the derivative can in principle be hedged by taking a short position in the same bonds. It is true that selling bonds short often entails additional costs, but this is not always so if existing long positions can be reduced, and in any case it does not seem defensible to include such costs in accounting fair value. The present proposal may present challenges if applied to derivatives management, since the bond-CDS bases are frequently difficult to estimate, introduce additional volatility in the fair value, and can become negative, producing counterintuitive results such as negative CFVA. However, if such issues are considered problematic, a different framework can be used for determining management P\&L and other performance metrics\footnote{In our view, a CVA desk should only be affected by risks that it is expected to manage. If it suffers from, for example, DVA volatility, it will have incentives to hedge DVA, while such hedging might not be desirable from a global balance sheet perspective: it introduces additional costs and risks with no clear associated benefit, while an unhedged DVA acts as a natural hedge for global earnings volatility.}.

Our ultimate goal is to be \textbf{compliant with IFRS 13}, which defines fair value as a non-entity-specific, market based, \textbf{exit price}. In turn, exit price is defined as the price obtained or paid when transferring an asset or liability, respectively, with an emphasis on the valuations performed by other market participants. The problem is that, strictly speaking, a derivative cannot be transferred. Its very nature depends on the two counterparties involved, and a transferred derivative is no longer the same derivative, exhibiting a different distribution of cash flows (since defaults alter the cash flow structure). Also, the bank can not freely transfer its side of the derivative to a different market participant. Instead the counterparty must agree to cancel the deal, and enter a new one with the market participant.

One solution then is to define exit price as the counterparty's replacement cost, shifting the issue of transfer to the counterparty. For an asymmetric derivative price, such as what is obtained with many approaches to FVA, this becomes awkward, resulting in an exit price depending on the counterparty's funding spread, but not the bank's, unless it is assumed that the third party involved (which takes over the bank's side of the derivative) has an identical funding spread as the bank. A symmetric formulation, such as the one defined here, does not have this problem, since price agreement is, in principal, always possible.

Furthermore, our formulation is \textbf{not entity-specific}. It is true that the proposed fair value depends on the bank's bond-CDS basis, but this is fundamentally not different from the dependence of DVA on its credit spreads. What we are doing is simply extending the credit component already present in DVA to encompass the full bond price, which is fixed by the market. What IFRS 13 actually means by a non-entity-specific valuation is that the exit price should not depend on such factors as the intention to hold the deal to maturity, etc. Another way of looking at this issue is to recognize that we are decomposing a given trade into a perfectly collateralized one, some bonds, and possibly CDS, all of which have well defined market values, determined by market participants independent of the entity.

A similar FVA term compliant with the existing accounting framework has also been obtained in \cite{Lou}, where a bilateral funding valuation adjustment is achieved, also depending on the bond-CDS bases of both counterparties. The point of departure, however, is rather different from the one shown here. While in this paper the value of an uncollateralized derivative has been decomposed into a perfectly collateralized portfolio plus bonds issued by both participants, finally leading to terms \ref{eq:CFVA} and \ref{eq:DFVA}, in \cite{Lou} it is argued that counterparties agreeing on a derivative transaction should be indifferent between entering its uncollateralized version or the collateralized equivalent with the collateral account accruing at the  financing rates of the corresponding participant to whom the derivative represents a liability. Such an assumption leads to non-recursive versions of CFVA and DFVA, that can be regarded as first-order approximations to the ones obtained here. Still, it is interesting that a result similar in spirit to ours follows from rather different assumptions.

It might seem confusing that so many alternative formulations for FVA exist. We have explained the shortcomings of alternate approaches, each one derived under varying assumptions. For instance, if we compare the approach taken here, with the analysis we performed in \cite{GunFer2014}, in the latter a deficit of collateral was funded by the bank itself, introducing a dependence on the bank's own bond-CDS spread for provided liquidity, while in this paper we recognize that such liquidity has a well-defined market value expressed in terms of the counterparty's bonds. 

A cost-based point of view underlies most approaches to FVA. It is argued that if \textbf{future costs} associated to a given derivatives transaction are not included in the valuation, the derivative will almost certainly lead to a loss over time. There are two problems with this statement. Firstly, the costs may not have been correctly estimated, such as is the case with approaches where a full FVA, based on the bank's full funding spread, is added to the CVA\footnote{As explained in Section \ref{sec:CVAplusFullFVA}, this does not take into consideration the \textbf{overlap between CVA and the bank's funding costs}. Trading with counterparties with a better credit quality than the bank itself will produce a funding benefit over time, and in the case of a balance sheet in equilibrium (with new deals replacing similar old deals on a continuous basis) this funding benefit is already incorporated into the current funding spread.}. Secondly, such an approach will behave in unpredictable ways when deals are terminated early. A counterparty should be charged upfront for any future estimated costs, by increasing the gross margin to anticipate them, but it would be paradoxical to charge the counterparty even more when closing out the deal, citing variations in expected future costs.

To illustrate why this could happen let us consider, tongue-in-cheek,  that besides an FVA, a manager with foresight decides to include an EVA, an Electricity Valuation Adjustment, in his derivative pricing. The operation of the derivatives desk will contribute to the electricity consumption of the bank, introducing a dependence on the electric spot price, he argues. Fortunately it can be hedged by entering into an electricity swap, and the cost of hedging can easily be included in the valuation. Counterparties start paying upfront in accordance with this new pricing, and time passes. Some time later, electricity prices suddenly plummet, but due to the hedges there is no impact in P\& L, and all is well. Until one day, when one of the largest counterparties asks to undo a large position with the bank.
\newline \noindent -'Alright', the bank manager says, 'but you will have to compensate me for the variation in my future electricity costs.'
\newline -'What? Electricity prices have fallen! And why should I have to compensate you for costs that you will not end up incurring, once the transaction has been closed?'
\newline -'I have hedged my electricity exposure, so closing out my hedges will imply an additional cost.'
\newline \noindent The bank manager continued insisting, and the counterparty had to accept the proposed close-out, but was naturally not pleased.
 
Next day, upper management came to see the manager. 
\newline \noindent -'We're afraid you're going to have to liquidate the portfolio we have with XYZ-bank. Capital ratios are looking a bit shaky, and we've decided that the position has to go.'
\newline \noindent Five minutes later, the manager had his contact at XYZ-bank on the phone.
\newline \noindent -'Sure! We'll terminate the position. You'll have to pay us for the close-out of our electricity hedge, though.', XYZ-bank said.
\newline -'Umm, I was kind of hoping that you would pay us...', the manager answered.
\newline -'Now, why would we do that?'
\newline \noindent In the end, the manager's desk took a loss corresponding to approximately twice the cost of closing out the EVA hedge.

The moral of this tale is that cost-based accounting only makes sense if assets and liabilities are guaranteed to be held to maturity. IFRS 13 explicitly forbids adopting this assumption, for good reasons. In a situation of stress a bank may be forced to liquidate large portions of its derivatives portfolio and will not have the luxury to affect the liquidation amounts, thereby running the risk of large losses. This would also exclude valuation adjustments, put forth recently in the literature, such as KVA (Capital Valuation Adjustment)\cite{GreenKenyon2014}. An exit price should be obtained from the counterparty's perspective, and will only coincide with the bank's viewpoint if, as we do, a symmetric approach is adopted.

\section*{Acknowledgements}


We would like to thank Juan Antonio de Juan Herrero, Jos\'e Manuel L\'opez P\'erez and Robert Dargavel Smith for helpful comments and suggestions.

\newpage

\appendix

\section{A formal derivation of the pricing formula}

\label{app:DerivationOfPricingEquation}

Let us now derive the pricing formula shown above with some analytical rigor. To do so, we will study a derivatives transaction and its replication building upon the setup described in \cite{Garcia2013} and \cite{GunFer2014}, but with the important new ingredient being the adoption of the Credit- and Debit Constraints \eqref{eq:CreditConstraint} and \eqref{eq:DebitConstraint}, respectively, which enforce the consistency of derivative- and bond valuations. The setup has the properties\footnote{These conditions can be easily relaxed, which would, with the exeption of the close-out assumptions, not affect the final result.}:

\begin{itemize}
\item The derivative depends on a single underlying market factor $S_t$ and has only a single payoff at maturity date $T$.
\item The parties involved in the transaction are separated into a Bank (B) and a Counterparty (C). The sign of its value is as seen by the bank.
\item Credit spreads are allowed to be stochastic.
\item The portfolio should be self-financing, with no additional funding obtained.
\item We assume a riskless close-out in the event that either bank or counterparty defaults, implying that claims put forth by either the surviving party or the liquidators of the defaulted party will be based upon the MtM of the corresponding perfectly collateralized transaction. No further defaults are thus considered when determining the close-out amounts.
\item Risk factors are, for simplicity, taken to be driven by single factor models, such as a single credit spread factor, and interest rates are deterministic.
\end{itemize}

In this setup, the bank will construct a replicating portfolio in a similar fashion as in \cite{Garcia2013} and \cite{GunFer2014}. We will skip some steps in the derivation of the strategy. The interested reader is referred to these works for more detail. We must note that the approach followed here will be that of replication, which is not necessarily the same thing as hedging. Although the difference between both concepts may appear subtle, it does plays a role in this context. Let us comment a bit on this before continuing with the derivation.

We use the term 'replication' in the sense of decomposition, so that the price of a given deal is simply equated to the price of a portfolio of simpler instruments (the replicating portfolio) which in aggregate produce the same cashflows and sensitivities to market variables as the original deal. By contrast, pricing by hedging would entail assigning a value to a given derivative based on the hedging transactions that the corresponding bank or financial institution is able (or willing) to perform. The rationale behind this approach states that a bank who intends to produce and sell a derivative will hedge its position in the market. The price to be charged to the customer, that could be regarded as the value of the transaction, would be that of the hedging portfolio. 

The limitation of the pricing-by-hedging viewpoint is that it cannot account for unhedgeable risks. As an example, a long position in a liquid bond has a well defined market price, but it is impossible to reproduce through hedging. The reason is that a hedge would entail shorting the bond, which would first have to be obtained via a repo transaction, with the ensuing reception of a repo-rate, leading to the bond yield not being fully recovered (apart from the fact that the cash lent in the repo would have to be funded, introducing a dependency on the hedger's funding spread). In the same way, when pricing derivatives by hedging, the bond-like characteristics of the derivative are simply discarded. Thus, we will define the replicating portfolio in terms of instruments with well-defined market value and exit-price, but which may, due to the presence of bonds, not be directly interpretable as a hedging portfolio. Finally, accounting considerations make it difficult to defend hedging approaches\footnote{As an example, IFRS 13 states that the fair value of a liability must reflect non-performance risk, that is, "the risk that the entity will not fulfill an obligation". Such a statement gives rise to the usual DVA term in annual reports of financial institutions. However, DVA is difficult to motivate in terms of hedging since no institution is able to completely hedge its own credit risk, including the jump-to-default component. But this component does exist and, as shown in \cite{GunFer2014}, has a well-defined value even for the institution considered as a going concern. Furthermore, IFRS 13 emphasizes that fair value must be a market-based measurement, and "not an entity-specific measurement", ie, the value of a derivative should not depend on internal decisions taken by the entity, such as the intention to hold the asset, to settle or to partially or totally hedge it.}.

The replicating portfolio will consist of a collateralized derivative $H_t$, used for capturing market risk, bonds issued by both the counterparty and the bank, a short-term CDS\footnote{The short-term (infinitesimal) CDSs are a theoretical construct (see \cite{Garcia2013} for more details) used here for simplicity, but in practice, the same hedging can be carried out by trading in two contracts of finite, but different, maturity.} written on the Counterparty, $CDS^C(t,\,t+dt)$, for eliminating counterparty's jump-to-default risk, plus another short-term CDS written on the Bank, $CDS^B(t,\,t+dt)$, for eliminating bank's jump-to-default risk\footnote{Although the bank does not have free access to its own CDS, as shown in \cite{GunFer2014}, this instrument has a well-defined value for it and lies implicitly in the derivative. Therefore, by including it in the replication portfolio, we are simply expliciting an existing component of the payoff. Alternatively, as in \cite{BurgardKjaer2012}, two bonds of different seniority could have been used.}.

Moreover, since credit spreads are taken to be stochastic, the bank will also need to eliminate these sources of risk. To do so, it will trade in bonds issued by the counterparty and itself, taking care that the portfolio remains self-financing, which can achieved in the following way:

\begin{itemize}
\item To replicate its own credit spread risk, the bank will be trading in bonds of its own of different maturities satisfying the debit constraint \eqref{eq:DebitConstraint}. Since here we will only be concerned with a single-factor model, just two bonds will be sufficient. For simplicity, we will take one of them, $B^B(t,\,t+dt)$, to be infinitesimally short-termed, while the other, $B^B(t,\,T)$, is of finite maturity. In this case, the debit constraint takes the form

\begin{equation}
-V_t^{-} = -\max\{-V_t,0\} = \Omega_t^B B^B(t,\,t+dt) + \omega_t^B B^B(t,\, T) \ ,
\label{eq:ConstraintBank}
\end{equation}
where $\Omega^B_t$ and $\omega^B_t$ are the quantities held of each bond. The bank could hedge this component by buying back its own debt of an amount precisely equal to $-V_t^{-}$.

\item To eliminate the counterparty's credit spread risk, the bank will trade in bonds issued by the counterparty satisfying the credit constraint \eqref{eq:CreditConstraint}. As before, we will use just two bonds, $B^C(t,\,t+dt)$ and $B^C(t,\,T)$. The credit constraint becomes:

\begin{equation}
V_t^{+} = \max\{V_t,0\} = \Omega_t^C B^C(t,\,t+dt) + \omega_t^C B^C(t,\, T) \ ,
\label{eq:ConstraintCount}
\end{equation}
with $\Omega^C_t$ and $\omega^C_t$ defined analogously as $\Omega^B_t$ and $\omega^B_t$.
\end{itemize}

Finally, we will have some cash in a collateral account, described using a unit-of-account $C_t$, of constant value 1, which generates an annualized interest of $c_t$, given by precisely the rate paid on collateral.

Putting all pieces together, the replicating portfolio will be

\begin{equation}
\begin{array}{ll}
V_t = & \alpha_t H_t + \beta _t C_t + \epsilon_t CDS^C(t,\,t+dt) + \eta_t CDS^B(t,\,t+dt)\\
& + \Omega_t^C B^C(t,\,t+dt) + \omega_t^C B^C(t,\, T) + \Omega_t^B B^B(t,\,t+dt) + \omega_t^B B^B(t,\, T) \ , \\ 
\end{array} 
\label{eq:ReplicatingPortfolio}
\end{equation}
where the Greek letters represent the amounts held of each instrument in the portfolio.

We assume that the evolution of the relevant market variables under the real measure $\mathbb{P}$ is described as

\begin{equation}
\left\lbrace
\begin{array}{l}
dS_t=\mu_{t}^{S} S_{t}dt+\sigma_{t}^{S}dW_{t}^{S,\mathbb{P}}\\
d\pi_t^{C}=\mu_t^{C}dt+\sigma_{t}^{C} dW_{t}^{C,\mathbb{P}}\\
d\pi_t^{B}=\mu_t^{B}dt+\sigma_{t}^{B} dW_{t}^{B,\mathbb{P}}\\
\end{array}
\right.
\label{eq:PDynamics}
\end{equation}
where $S_t$ represents the price of the derivative's underlying asset at time $t$, while $\pi_t^C$ and $\pi_t^B$ are the short term CDS spread of the counterparty and bank, respectively. These spreads are defined so that $CDS^{k}(t,t+dt)=0, k\in\{C,B\}$. $\mu_{t}^{S}$, $\mu_{t}^{C}$ and $\mu_{t}^{B}$ are the real world drifts of these processes, while $\sigma_{t}^{S}(t,S_{t})$, $\sigma_{t}^{C}(t,\pi_{t}^{C})$, $\sigma_{t}^{B}(t,\pi_{t}^{B})$ are their volatilities. Interest rates will be taken to be deterministic.

The three processes will be correlated with time dependent correlations:
\begin{equation}
\begin{array}{ccc}
\rho_{t}^{S,C}dt=dW_{t}^{S,\mathbb{P}} dW_{t}^{C,\mathbb{P}}&\rho_{t}^{B,C}dt=dW_{t}^{B,\mathbb{P}} dW_{t}^{C,\mathbb{P}}&\rho_{t}^{S,B}dt=dW_{t}^{S,\mathbb{P}} dW_{t}^{B,\mathbb{P}}\\
\end{array}
\end{equation}

Two additional sources of uncertainty are described by the default indicator processes $N_{t}^{C,\mathbb{P}}=1_{\{\tau^{C}\leq t\}}$ and $N_{t}^{B,\mathbb{P}}=1_{\{\tau^{B}\leq t\}}$, with real world intensities $\lambda_{t}^{C,\mathbb{P}}$ and $\lambda_{t}^{B,\mathbb{P}}$, with $\tau^{C}$ and $\tau^B$ being the default times of the counterparty and the bank, respectively.

To do the replication, we will proceed in the standard way, equating the differential of \eqref{eq:ReplicatingPortfolio}, assuming a self-financing strategy, with the expression obtained by expanding $dV_t$ using It\^o's Lemma, and choosing the available coefficients so that the stochastic terms cancel. The remaining, deterministic terms then imply a differential equation for $V_t$. Since this procedure is fairly standard, for the sake of brevity, we will omit some steps, and not spell out explicitly certain terms.

Conditional on both the counterparty and the bank being alive at time $t$, the change in $V_t$ will be given by (applying It\^o's Lemma for jump diffusion processes)

\begin{equation}
\begin{array}{ll}
dV_{t}=&\mathcal{L}_{SCB}V_{t}dt+\frac{\partial V_t}{\partial S_t}S_{t}\sigma_{t}^{S}dW_{t}^{S,\mathbb{P}}+\frac{\partial V_t}{\partial \pi_{t}^{C}} \sigma_{t}^{C}dW_{t}^{C,\mathbb{P}}+\frac{\partial V_t}{\partial \pi_{t}^{B}} \sigma_{t}^{B}dW_{t}^{B,\mathbb{P}}\\
&+\Delta V_{t}^{C}dN_{t}^{C,\mathbb{P}}+\Delta V_{t}^{B}dN_{t}^{B,\mathbb{P}},\\
\end{array}
\label{eq:dVFromIto}
\end{equation}
where $\mathcal{L}_{SCB}V_{t}$ groups together deterministic terms.

On the other hand, by assuming a self-replicating trading strategy, taking the differential of \eqref{eq:ReplicatingPortfolio} gives

\begin{equation}
\begin{array}{ll}
dV_t = & \alpha_t dH_t + \beta _t dC_t + \epsilon_t dCDS^C(t,\,t+dt) + \eta_t dCDS^B(t,\,t+dt)\\
& + \Omega_t^C dB^C(t,\,t+dt) + \omega_t^C dB^C(t,\, T) + \Omega_t^B dB^B(t,\,t+dt) + \omega_t^B dB^B(t,\, T)\\
\end{array}
\label{eq:DifferentialReplicatingPortfolio}
\end{equation}

We write down the differentials of the short- and long term bonds, including terms corresponding to their jump-to-default, as well as the instantaneous CDS and the collateral account\footnote{Actually, as explained in \cite{BrigoBuescu2012}, this formulation of the self-financing condition is an abuse of notation. For example, we have stated that $C_t$ is a unit-of-account of constant value 1, and should therefore obey $dC_t = 0$. However, it should be understood implicitly, when reading \eqref{eq:DifferentialReplicatingPortfolio}, that the differentials of the different components refer to the \textit{gain processes}, including generated dividends.}:

\begin{equation}
\left\lbrace
\begin{array}{l}
dB^k(t,t+dt)=f_{t}^{k}B^{k}(t,t+dt)dt + (R_{k}-1)B^{k}(t,t+dt)dN_{t}^{k,\mathbb{P}}\\
dB^k(t,T)=\mathcal{L}_{k}B^{k}(t,T)dt+\frac{\partial B^{k}(t,T)}{\partial \pi_{t}^{k}} \sigma_{t}^{k}dW_{t}^{k,\mathbb{P}}+\Delta B^{k}(t,T)dN_{t}^{k,\mathbb{P}}\\
dCDS^{k}(t,t+dt)=\pi_{t}^{k}dt-(1-R_k)dN_{t}^{k,\mathbb{P}}\\
dC_t=c_tdt\\
\end{array}
\right.
\label{eq:BondDynam}
\end{equation}
for $k\in\{C,B\}$, where $f_{t}^{k}=c_t+\bar{f}_{t}^{k}$ represents the short term funding rate, $\bar{f}_{t}^{k}$ is the short term funding spread over the OIS rate $c_t$, $R_k$ is the recovery rate upon default, and

\begin{equation*}
\begin{array}{c}
\mathcal{L}_{k}B^{k}(t,T)=\frac{\partial B^{k}(t,T)}{\partial t}+\mu_{t}^{k}\frac{\partial B^{k}(t,T)}{\partial \pi_t^{k}}+\frac{1}{2}(\sigma_t^{k})^{2}\frac{\partial^2 B^{k}(t,T)}{\partial^2 \pi_t^{k}} \\
\end{array}
\end{equation*}

If we substitute all the differential terms in (\ref{eq:DifferentialReplicatingPortfolio}), we reach a replicating equation in which we eliminate the stochastic terms driven by $dW_{t}^{k,\mathbb{P}}$, $k\in\{C,B,S\}$ and $dN_{t}^{C,\mathbb{P}}$, $dN_{t}^{B,\mathbb{P}}$ by taking

\begin{equation}
\begin{array}{cc}
\alpha_{t}=\frac{\frac{\partial V_{t}}{\partial S_{t}}}{\frac{\partial H_{t}}{\partial S_{t}}}& 
\omega_{t}^C=\frac{\frac{\partial V_{t}}{\partial \pi_{t}^{C}}}{\frac{\partial B^C(t,T)}{\partial \pi_{t}^{C}}}\\
\omega_{t}^B=\frac{\frac{\partial V_{t}}{\partial \pi_{t}^{B}}}{\frac{\partial B^B(t,T)}{\partial \pi_{t}^{B}}}&\epsilon_{t}=-V_t^{+}-\frac{\Delta V_{t}^{C}}{1-R_C}\\
\multicolumn{2}{c}{\eta_t = V_t^{-} - \frac{\Delta V_{t}^{B}}{1-R_B}}\\
\end{array}
\label{eq:HedgWeights}
\end{equation}
while $\Omega^k_t$, $k\in\{C,B\}$, are fixed by the constraints \eqref{eq:ConstraintBank} and \eqref{eq:ConstraintCount}.

Making use of the PDEs for $H_t$, $B^C(t,T)$ and $B^B(t,T)$, we obtain the final PDE:

\begin{equation}
\hat{\mathcal{L}}_{SCB}V_{t} +\frac{\pi_{t}^{C}}{1-R_C}\Delta V_{t}^{C}+\frac{\pi_{t}^{B}}{1-R_B}\Delta V_{t}^{B}=(f_t^{C}-\pi_{t}^{C})V_{t}^{+}-(f_t^{B}-\pi_{t}^{B})V_{t}^{-} \ ,
\label{eq:FinalPDE}
\end{equation}
where

\begin{equation*}
\begin{array}{ll}
\hat{\mathcal{L}}_{SCB}V_{t}=&\frac{\partial V_t}{\partial t}+(r_t-q_t)S_t\frac{\partial V_t}{\partial S_t} + (\mu_t^B-M_t^B\sigma_t^B)\frac{\partial V_t}{\partial \pi_t^B}+ (\mu_t^C-M_t^C\sigma_t^C)\frac{\partial V_t}{\partial \pi_t^C}\\
&+\frac{1}{2}\frac{\partial^2 V_t}{\partial S_t^2}S_t^2(\sigma_t^S)^2+\frac{1}{2}\frac{\partial^2 V_t}{\partial \pi_t^{B^2}}(\sigma_t^B)^2+ \frac{1}{2}\frac{\partial^2 V_t}{\partial \pi_t^{C^2}}(\sigma_t^C)^2\\
&+\frac{\partial^2 V_t}{\partial S_t\partial \pi_t^B}S_t\sigma_t^S\sigma_t^B\rho_t^{S,B}+\frac{\partial^2 V_t}{\partial S_t\partial \pi_t^C}S_t\sigma_t^S\sigma_t^C\rho_t^{S,C}+\frac{\partial^2 V_t}{\partial \pi_t^C\partial \pi_t^B} \sigma_t^C\sigma_t^B\rho_t^{C,B} \ ,
\end{array}
\end{equation*}
$M_t^C$ and $M_t^B$ are the market price of credit risk of the counterparty and bank, respectively, that is, the expected excess return of a credit derivative on each of them over the collateral rate divided by the derivatives' volatility.

Once we have arrived to the PDE depicted in (\ref{eq:FinalPDE}), we can follow the well-known steps leading to the Feynman-Kac formula (shown in, for example, \cite{KarShreve}). To do so, we define the process:
\begin{equation}
X_t=V_t\exp\Big(-\int_{s=0}^{t}c_s\,ds\Big)1_{\{\tau^C>t\}}1_{\{\tau^B>t\}}
\label{eq: ProcessX}
\end{equation}

We then place ourselves in the risk-neutral measure $\mathbb{Q}$ in which the drifts of $S_t$, $\pi_t^B$ and $\pi_t^C$ are given by $(r_t-q_t)S_t$, $\mu_t^B-M_t^B\sigma_t^B$ and $\mu_t^C-M_t^C\sigma_t^C$, respectively. Furthermore, default intensities of the counterparty and bank are given by:
\begin{equation*}
\begin{array}{ll}
\lambda_t^{C,\mathbb{Q}}=\frac{\pi_t^C}{1-R_C}& \lambda_t^{B,\mathbb{Q}}=\frac{\pi_t^B}{1-R_B}\\
\end{array}
\end{equation*}
We apply It\^o's Lemma for jump diffusion processes to $X_t$ in $\mathbb{Q}$, using the stochastic discount factor notation already introduced in \eqref{eq:StochasticDiscountFactor}
\begin{equation*}
\begin{array}{ll}
dX_t=&D(0,t)\Big[1_{\{\tau^C>t\}}1_{\{\tau^B>t\}}\Big(-c_tV_tdt+\hat{\mathcal{L}}_{SCB}V_{t} dt+\frac{\partial V_t}{\partial S_t}S_t\sigma_t^SdW_t^S\\
&+\frac{\partial V_t}{\partial \pi_t^C}\sigma_t^CdW_t^C+\frac{\partial V_t}{\partial \pi_t^B}\sigma_t^BdW_t^B\Big)-1_{\{\tau^C>t\}}V_tdN_t^{B,\mathbb{Q}}-1_{\{\tau^B>t\}}V_tdN_t^{C,\mathbb{Q}}\Big] \ ,\\
\end{array}
\end{equation*}
while from the PDE shown above, we have:

\begin{equation*}
\begin{array}{ll}
\hat{\mathcal{L}}_{SCB}V_{t}&=(f_t^{C}-\pi_{t}^{C})V_{t}^{+}-(f_t^{B}-\pi_{t}^{B})V_{t}^{-}-\lambda_t^{C,\mathbb{Q}}\Delta V_{t}^{C}-\lambda_t^{B,\mathbb{Q}}\Delta V_{t}^{B}\\
&=(c_t+\bar{f}_t^{C}-\pi_{t}^{C})V_{t}^{+}-(c_t+\bar{f}_t^{B}-\pi_{t}^{B})V_{t}^{-}-\lambda_t^{C,\mathbb{Q}}\Delta V_{t}^{C}-\lambda_t^{B,\mathbb{Q}}\Delta V_{t}^{B}\\
&=c_tV_t+(\bar{f}_t^{C}-\pi_{t}^{C})V_{t}^{+}-(\bar{f}_t^{B}-\pi_{t}^{B})V_{t}^{-}-\lambda_t^{C,\mathbb{Q}}\Delta V_{t}^{C}-\lambda_t^{B,\mathbb{Q}}\Delta V_{t}^{B}\\
\end{array}
\end{equation*}
so that, naming $\gamma_t^k=\bar{f}_t^k - \pi_t^k$, $k\in\{C,B\}$, we get

\begin{equation*}
\begin{array}{ll}
dX_t=&D(0,t)\Big[1_{\{\tau^C>t\}}1_{\{\tau^B>t\}}\Big((\bar{f}_t^{C}-\pi_{t}^{C})V_{t}^{+}dt-(\bar{f}_t^{B}-\pi_{t}^{B})V_{t}^{-}dt\\
&-\lambda_t^{C,\mathbb{Q}}\Delta V_{t}^{C}dt-\lambda_t^{B,\mathbb{Q}}\Delta V_{t}^{B}dt+\frac{\partial V_t}{\partial S_t}S_t\sigma_t^SdW_t^S\\
&+\frac{\partial V_t}{\partial \pi_t^C}\sigma_t^CdW_t^C+\frac{\partial V_t}{\partial \pi_t^B}\sigma_t^BdW_t^B\Big)-1_{\{\tau^C>t\}}V_tdN_t^{B,\mathbb{Q}}-1_{\{\tau^B>t\}}V_tdN_t^{C,\mathbb{Q}}\Big] \ ,\\
\end{array}
\end{equation*}

Next, we can integrate between $t$ and $T$ and take expectations conditional on $\mathcal{F}_t$. Terms in $dW$ disappear since they are expected values of It\^o integrals. Furthermore, terms in $\lambda$ also vanish due to the definition of default intensity, allowing us to write:

\begin{equation}
\begin{array}{ll}
V_t=&E_\mathbb{Q}\Big[V_T D(t,T) 1_{\{\tau^C>T\}}1_{\{\tau^B>T\}}|\mathcal{F}_t\Big]\\
&+E_\mathbb{Q}\Big[\int_{s=t}^{T}1_{\{\tau^B>s\}}D(t,s)(V_s+\Delta V_{s}^{C})dN_s^{C,\mathbb{Q}}|\mathcal{F}_t\Big]\\
&+E_\mathbb{Q}\Big[\int_{s=t}^{T}1_{\{\tau^C>s\}}D(t,s)(V_s+\Delta V_{s}^{B})dN_s^{B,\mathbb{Q}}|\mathcal{F}_t\Big]\\
&-E_\mathbb{Q}\Big[\int_{s=t}^{T}1_{\{\tau^C>s\}}1_{\{\tau^B>s\}}D(t,s)\gamma_s^{C}V_{s}^{+}ds|\mathcal{F}_t\Big]\\
&+E_\mathbb{Q}\Big[\int_{s=t}^{T}1_{\{\tau^C>s\}}1_{\{\tau^B>s\}}D(t,s)\gamma_s^{B}V_{s}^{-}ds|\mathcal{F}_t\Big]\\
\end{array}
\label{eq:FirstExpect}
\end{equation}

We let $V_t^c$ be the time $t$ value that the derivative would have if it were perfectly collateralized. Now, if we assume, in accordance with a riskless close-out, that after the counterparty's default, $V_s$ jumps to $R_CV_s^c$ if $V_s^c \geq 0$ and to $V_s^c$ if $V_s^c < 0$, and that after the bank's default $V_s$ jumps to $V_s^c$ if $V_s^c \geq 0$ and to $R_BV_s^c$ if $V_s^c< 0$, then, after some indicator manipulations (see, for example, Appendix A of \cite{GunFer2014} for details), we arrive at

\begin{equation}
\begin{array}{ll}
V_t=&E_\mathbb{Q}\Big[V_T D(t,T)|\mathcal{F}_t\Big]\\
&-E_\mathbb{Q}\Big[\int_{s=t}^{T}\int_{u=s}^{+\infty}D(t,s)(1-R_C)(V_s^c)^{+}dN_u^{B,\mathbb{Q}}dN_s^{C,\mathbb{Q}}|\mathcal{F}_t\Big]\\
&+E_\mathbb{Q}\Big[\int_{s=t}^{T}\int_{u=s}^{+\infty}D(t,s)(1-R_B)(V_s^c)^{-}dN_u^{C,\mathbb{Q}}dN_s^{B,\mathbb{Q}}|\mathcal{F}_t\Big]\\
&-E_\mathbb{Q}\Big[\int_{s=t}^{T}1_{\{\tau^C>s\}}1_{\{\tau^B>s\}}D(t,s)\gamma_s^{C}V_{s}^{+}ds|\mathcal{F}_t\Big]\\
&+E_\mathbb{Q}\Big[\int_{s=t}^{T}1_{\{\tau^C>s\}}1_{\{\tau^B>s\}}D(t,s)\gamma_s^{B}V_{s}^{-}ds|\mathcal{F}_t\Big]\\
\end{array}
\label{eq:FinalPricingEquationForm3}
\end{equation}

Intermediate cash flows can easily be incorporated by substituting the first term on the RHS of \eqref{eq:FinalPricingEquationForm3} for the complete perfectly collateralized price $V_t^c$, which discounts all cash flows to present time using OIS rate $c_t$.

\section{Practical approximations}
\label{app:AFirstOrder}

The obtained pricing equation \eqref{eq:FinalPricingEquationForm3} may be difficult to evaluate directly, due to its recursive nature. For this reason we will now provide a couple of approximations that can be used in practice. We begin by suggesting the omission of the recursivity in the CFVA and DFVA terms,  providing a straightforward option. We then proceed to study whether it would be valid to drop both FVA terms and instead calculate contingent debit and credit valuation adjustments based on the default probabilities extracted from bonds issued by the bank and its counterparty, respectively, and thereby incorporate funding considerations via the CVA and DVA terms. As we will see, the resulting expression equals the pricing equation to a first-order approximation.

Let us first rewrite \eqref{eq:FinalPricingEquationForm3} using the definition of default intensity.

\begin{equation}
\begin{array}{ll}
V_t=&E_\mathbb{Q}\Big[V_T D(t,T)|\mathcal{F}_t\Big]\\
&-E_\mathbb{Q}\Big[\int_{s=t}^{T}\lambda_s^C1_{\{\tau^C>s\}}1_{\{\tau^B>s\}}D(t,s)(1-R_C)(V_s^c)^{+}ds|\mathcal{F}_t\Big]\\
&+E_\mathbb{Q}\Big[\int_{s=t}^{T}\lambda_s^B1_{\{\tau^C>s\}}1_{\{\tau^B>s\}}D(t,s)(1-R_B)(V_s^c)^{-}ds|\mathcal{F}_t\Big]\\
&-E_\mathbb{Q}\Big[\int_{s=t}^{T}1_{\{\tau^C>s\}}1_{\{\tau^B>s\}}D(t,s)\gamma_s^{C}V_{s}^{+}ds|\mathcal{F}_t\Big]\\
&+E_\mathbb{Q}\Big[\int_{s=t}^{T}1_{\{\tau^C>s\}}1_{\{\tau^B>s\}}D(t,s)\gamma_s^{B}V_{s}^{-}ds|\mathcal{F}_t\Big]\\
\end{array}
\label{eq:FinalPricingEquationForm4}
\end{equation}

If we iterate the integral equation, the future derivative exposures $V_s^\pm$ entering into the FVA terms become $\left(V_s^c - CVA_s+DVA_s-CFVA_s + DFVA_s \right)^\pm $. Since the FVA terms will be $\mathcal{O}(\gamma ^2)$, they can be dropped in a first-order approximation in the bond-CDS bases (which tend to be small). Also, in most cases $CVA_s$ and $DVA_s$ will be small compared to $V_s^c$, and we can drop their feedback into the FVA terms as well. The upshot is that in the FVA terms the full exposure $V_s$ can be approximated by its collateralized equivalent $V_s^c$. Recalling that $\pi_s^k=\lambda_s^k(1-R_k)$, $k\in\{B,C\}$ we therefore get

\begin{equation}
\begin{array}{ll}
V_t=&E_\mathbb{Q}\Big[V_T D(t,T)|\mathcal{F}_t\Big]\\
&-E_\mathbb{Q}\Big[\int_{s=t}^{T}1_{\{\tau^C>s\}}1_{\{\tau^B>s\}}D(t,s)(\pi_s^C+\gamma_s^C)(V_s^c)^{+}ds|\mathcal{F}_t\Big]\\
&+E_\mathbb{Q}\Big[\int_{s=t}^{T}1_{\{\tau^C>s\}}1_{\{\tau^B>s\}}D(t,s)(\pi_s^B+\gamma_s^B)(V_s^c)^{-}ds|\mathcal{F}_t\Big]\\
\end{array}
\label{eq:CVAFCAFBA}
\end{equation}

We recall that $\pi_s^k+\gamma_s^k = \bar{f}_{s}^{k}$ is the short term funding spread over the OIS rate $c_t$. This equation is much simpler to evaluate than \eqref{eq:FinalPricingEquationForm4}, and in fact the second and third terms are structurally equivalent to the Funding Cost Adjustment (FCA) and Funding Benefit Adjustment (FBA) found in, for example, \cite{Gregory2012}. The difference is that in \cite{Gregory2012} the FCA term is governed by the bank's, and not the counterparty's, funding spread, and the overlap with CVA is also not corrected for. Here, a separate term for CVA is not obtained since it is fully contained in the FCA term. Keeping CVA and DVA separate, we thus have the approximate expressions
\begin{equation}
CFVA_t = E_\mathbb{Q}\Big[\int_{s=t}^{T}1_{\{\tau^C>s\}}1_{\{\tau^B>s\}}D(t,s)\gamma_s^C(V_s^c)^{+}ds|\mathcal{F}_t\Big] \ ,
\end{equation}
and
\begin{equation}
DFVA_t = E_\mathbb{Q}\Big[\int_{s=t}^{T}1_{\{\tau^C>s\}}1_{\{\tau^B>s\}}D(t,s)\gamma_s^B(V_s^c)^{-}ds|\mathcal{F}_t\Big] \ .
\end{equation}

Now, returning to equation \eqref{eq:CVAFCAFBA}: if we bootstrapped default probabilities from a bond, the funding spread $\bar{f}_{t}^{k}$ can be approximated as the instantaneous default probability extracted from the bond, $\bar{\lambda}_{t}^{k}$, times the loss-given-default\footnote{Imagine a one-period zero coupon bond that gave us 1 at time 1 with probability $1-p$ and $R$ with probability $p$, being $c$ the constant risk-free rate between $t=0$ and $t=1$. Using standard valuation theory, the price of this bond at time 0 is $B(0)=(1-p)e^{-c}+pRe^{-c}=e^{-c}(1-(1-R)p)$, which can be approximated as $B(0)=e^{-(c+(1-R)p)}$. Therefore, the funding spread needed to match the price of the bond equals its probability of default times $1-R$.}. Thus, the resulting equation is:

\begin{equation}
\begin{array}{ll}
V_t=&E_\mathbb{Q}\Big[V_T D(t,T)|\mathcal{F}_t\Big]\\
&-E_\mathbb{Q}\Big[\int_{s=t}^{T}1_{\{\tau^C>s\}}1_{\{\tau^B>s\}}D(t,s)\bar{\lambda}_{s}^{C}(1-R_C)(V_s^c)^{+}ds|\mathcal{F}_t\Big]\\
&+E_\mathbb{Q}\Big[\int_{s=t}^{T}1_{\{\tau^C>s\}}1_{\{\tau^B>s\}}D(t,s)\bar{\lambda}_{s}^{B}(1-R_B)(V_s^c)^{-}ds|\mathcal{F}_t\Big]\\
\end{array}
\end{equation}

Finally, the distributions of the default indicator functions $1_{\{\tau^k>s\}}$ depend on the default probabilities $\lambda_s^k=\bar{\lambda}_s^k - \gamma_s^k/(1-R_k) = \bar{\lambda}_s^k + \mathcal{O}(\gamma^k)$. Thus, at first-order approximation in $\gamma ^k$, and back to the notation of \eqref{eq:FinalPricingEquationForm3}, we obtain the following expression:

\begin{equation}
\begin{array}{ll}
V_t=&E_\mathbb{Q}\Big[V_T D(t,T)|\mathcal{F}_t\Big]\\
&-E_\mathbb{Q}\Big[\int_{s=t}^{T}\int_{u=s}^{+\infty}D(t,s)(1-R_C)(V_s^c)^{+}d\bar{N}_u^{B,\mathbb{Q}}d\bar{N}_s^{C,\mathbb{Q}}|\mathcal{F}_t\Big]\\
&+E_\mathbb{Q}\Big[\int_{s=t}^{T}\int_{u=s}^{+\infty}D(t,s)(1-R_B)(V_s^c)^{-}d\bar{N}_u^{C,\mathbb{Q}}d\bar{N}_s^{B,\mathbb{Q}}|\mathcal{F}_t\Big]\\
\end{array}
\end{equation}
where the $d\bar{N}_s^{k,\mathbb{Q}}$ are default processes inferred from the bond curves. We have thus also shown that the CFVA and DFVA terms can be absorbed, to first order, into the CVA and DVA adjustments. 

\section{Consistency with Bond Prices}

\label{app:BondConsistency}

We have constructed the pricing equation \eqref{eq:FinalPricingEquationForm2} taking into consideration bond-CDS bases in order to be consistent with both the counterparty's and the bank's bonds. But can we actually reproduce observed bond prices in the current framework? Yes, but we have to be careful when doing so. The difference between bonds and derivatives is that the latter will be liquidated upon default of either counterparty. A bond, by contrast, is traded in the market, and the market itself cannot in general cause the bond to be liquidated. So the first step in reproducing bond prices is to make the bond buyer default-free, based on the understanding that if the buyer of the bond were to default, the bond would simply be sold in the market with no change in value\footnote{An exception of course is if the buyer is a systemic entity, correlated with the bond issuer. Such systemic dependence should already be contained in the issuer's bond curve, however, and so when pricing we can assume that the buyer is a small, with no systemic impact}. 

Assuming that the bank is the buyer (the bond is an asset), the pricing equation for the bond would then be
\begin{equation}
V_t = V_t^c - CVA_t -CFVA_t \ ,
\label{eq:PricingEquationBond}
\end{equation}
where there is no DVA since the bank's default is irrelevant to the bond price, and no DFVA since the exposure is always positive for an asset. 

Equation \ref{eq:FinalPricingEquationForm3} then becomes
\begin{equation}
\begin{array}{ll}
V_t=&V_t^c-E_\mathbb{Q}\Big[\int_{s=t}^{T}D(t,s)(1-R_C)V_s^cdN_s^{C,\mathbb{Q}}|\mathcal{F}_t\Big]-E_\mathbb{Q}\Big[\int_{s=t}^{T}1_{\{\tau^C>s\}}D(t,s)\gamma_s^{C}V_{s}ds|\mathcal{F}_t\Big]\\
\end{array}
\label{eq:FinalPricingEquationBond}
\end{equation}
where positive exposures $V_s^{+}$ have been substituted for $V_s$ since the valuation will always be positive.

Let us now check what this equation gives for the simple case of a default-free issuer and a zero-coupon bond of notional $N$ and maturity $T$, and under the simplified assumption that the discount rate $c$, as well as the bond-CDS basis $\gamma$, are deterministic and constant. Then there will be no CVA, the deal will always be alive, and
$$CFVA_t =  \int_t^T D(t,\,s)\, \gamma \, V_s \, ds \ .$$
Using that $V_t^c = e^{-c(T-t)}N$ The full pricing equation will therefore be
$$V_t = e^{-c(T-t)}N - \int_t^T e^{-c(s-t)}\, \gamma  \, V_s \, ds \ .$$
It can easily be checked that the solution to this equation is
$$V_t = e^{-(c+\gamma)(T-t)} \ ,$$
and so we recover the bond price corresponding to a default-free bond.

The issue gets more complicated when credit risk is involved, and solving the equation does not directly produce a simple exponential expression for the bond price. The reason is subtle, and related to the different recovery conventions used when modeling the defaults of derivatives and bonds, respectively. For derivatives we are, as explained in Section \ref{sec:ReconcilingTwoWorlds}, assuming a riskless close-out, which means that the post-default value of a derivative is taken as a multiple $R$ of the riskless (or perfectly collateralized) value. By contrast, for mathematical simplicity we have modeled, in Appendix \ref{app:DerivationOfPricingEquation}, the recovery of the bonds in the replicating portfolio as relative, meaning that they denote the fraction of the full pre-default value that is recovered after default. We can easily accommodate such a recovery convention by performing the substitution 
$$R \rightarrow R\frac{V_t}{V_t^c}$$
in equation \eqref{eq:FinalPricingEquationBond}. By doing so, and solving the equation, one obtains the price of the zero coupon bond as
\begin{equation}
V_t = e^{-(c+\pi+\gamma)(T-t)}N \ ,
\label{eq:BondCurveValuation}
\end{equation}
where $\pi$ is the CDS spread (again assuming that all parameters are constant in time). The notional amount is therefore simply discounted by the bond curve.

Yet a third approach is the common standard consisting of defining the recovery fraction as the proportion of the notional amount that is recovered. In general, holding all other parameters constant, different bond prices will be obtained for different recovery conventions. Market prices are unique, though, which implies that bond-CDS bases directly obtained from bond curves will differ depending on the recovery convention adopted, in such a way as to compensate for the differences. The bond-CDS basis we use, is that of a relative recovery convention, so if such a convention is adopted, the bond-CDS basis can easily be obtained by inferring the bond curve used to discount bond cash flows. Other recovery conventions are not on the same footing as the relative one, but the difference in bond-CDS bases should not be large.

In order to be able to obtain greater precision, we will derive expressions for bond prices from \eqref{eq:FinalPricingEquationBond}, which can be used to calibrate the bond-CDS basis parameters $\gamma_t^X$, $X \in \{C,B\}$ to market prices. An alternative is of course to use observed bond-CDS bases directly, but it should be born in mind that strictly speaking the bases introduced in Appendix \ref{app:DerivationOfPricingEquation} correspond to bonds using a relative recovery convention. In Section \ref{app:BondRiskless} we derive the value for a bond under a riskless recovery assumption (the default value of the bond is the recovery fraction times the bond valuation obtained by discounting using the OIS curve), while we obtain the price corresponding to an absolute recovery convention (in which the default value is a constant recovery fraction of the notional) in Section \ref{app:BondAbsolute}.

\subsection{Bond with Riskless Recovery}

\label{app:BondRiskless}

Let us assume a bond issued by the counterparty with notional $N$, generating a series of cash flows $c_i$ at payment dates $I_c=\{t_i^{P}\}$. As stated above, the bond's holder can be assumed not to affect its price, so we can suppose that the bank cannot default, that is, $\lambda^{B}=0$, so we will only care for those variables related to the counterparty (the bank's bond-CDS basis is also irrelevant since $V_t^{-}=0$). For simplicity, we will assume that recoveries and interest rates (and therefore, discount factors) are deterministic. Furthermore, hazard rates $\lambda_j$ and bond-CDS basis $\gamma_k$ are deterministic and display a piecewise-constant structure, with parameters remaining constant between dates $\{t_j^{\lambda}\}$ and $\{t_k^{\gamma}\}$, respectively. Let us define $\{t_A\}=\{t_i^{P}\}\cup \{t_j^{\lambda}\} \cup \{t_k^{\gamma}\}$.

Let us first consider the case in which we only have one period: $[t_0,t_1]$. The bond only pays at date $t_1$ a cash flow equal to $c_1$ (which includes the notional).  With this in mind, substituting in \eqref{eq:FinalPricingEquationBond} for $t\in[t_0,t_1]$ yields:

\begin{equation}
\begin{array}{ll}
V_t&=D(t,t_1)c_1-\gamma_{1}\int_{s=t}^{t_1}e^{-\lambda_1(s-t)}D(t,s)V_sds\\
&-\int_{s=t}^{t_1}\lambda_1(1-R)D(s,t_1)c_1e^{-\lambda_1(s-t)}D(t,s)ds\\
\end{array}
\label{eq:OnePeriodBond}
\end{equation}
where we have used that $(V_{s}^{c})^{+}=D(s,t_1)c_1$. If we rewrite the expression, evaluate the latter integral and multiply everything by $D^{-1}(t,t_1)/c_1$, we get:

\begin{equation}
\begin{array}{ll}
D^{-1}(t,t_1)\frac{V_t}{c_1}&=1-\gamma_{1}\int_{s=t}^{t_1}e^{-\lambda_1(s-t)}D^{-1}(s,t_1)\frac{V_s}{c_1}ds-(1-R)(1-e^{-\lambda_1(t_1-t)})\\
\end{array}
\end{equation}

Let us define $V_t^{*}= D^{-1}(t,t_1)\frac{V_t}{c_1}$. Then we have

\begin{equation}
\begin{array}{ll}
V_t^{*}&=1-\gamma_{1}\int_{s=t}^{t_1}e^{-\lambda_1(s-t)}V_s^{*}ds-(1-R)(1-e^{-\lambda_1(t_1-t)})\\
\end{array}
\end{equation}

Performing a change of variables by defining $\tilde{t}=\lambda_1(t_1-t)$ and $\tilde{s}=\lambda_1(t_1-s)$, together with the notation $\tilde{V}_{\tilde{t}}=V_t^{*}$, we get

\begin{equation}
\begin{array}{ll}
\tilde{V}_{\tilde{t}}&=1-\frac{\gamma_{1}}{\lambda_1}e^{-\tilde{t}}\int_{\tilde{s}=0}^{\tilde{t}}e^{\tilde{s}}\tilde{V}_{\tilde{s}}ds-(1-R)(1-e^{-\tilde{t}})\\
\end{array}
\end{equation}

If we now multiply all terms by $e^{\tilde{t}}$ and differentiate with respect to $\tilde{t}$, we have transformed the integral equation into the ordinary differential equation

\begin{equation}
\begin{array}{l}
\tilde{V}_{\tilde{t}}^{'}+\Big(1+\frac{\gamma_1}{\lambda_1}\Big)\tilde{V}_{\tilde{t}}=R\\
\end{array}
\end{equation}

Solving this equation and undoing previous changes of variables, we find:

\begin{equation}
\begin{array}{l}
D^{-1}(t,t_1)\frac{V_t}{c_1}=\frac{R\lambda_1}{\lambda_1+\gamma_1}+\bar{K}e^{-(\lambda_1+\gamma_1)(t_1-t)}\\
\end{array}
\label{eq:SolvDifEq1}
\end{equation}
where $\bar{K}$ is an integration constant. Imposing the terminal condition $V_{t_1}=c_1$, gives

\begin{equation}
\begin{array}{ll}
V_t&=c_1D(t,t_1)\Big[1-\frac{(1-R)\lambda_1+\gamma_1}{\lambda_1+\gamma_1}\Big(1-e^{-(\lambda_1+\gamma_1)(t_1-t)}\Big)\Big]\\ 
\end{array}
\label{eq:Bond1FinalPrice}
\end{equation}

Next we consider the case with two periods: $[t_0,t_1]$ and $[t_1,t_2]$, with cash flows $c_1$ at $t_1$ and $c_2$ at $t_2$. First, we can restrict ourselves to the period $[t_1,t_2]$. There, we can apply the solution \eqref{eq:Bond1FinalPrice}. If we name $V_{t_{1}^{+}}$ as the value of the bond at date $t_1^{+}$, that is, immediately after cash flow $c_1$ at $t_1$ is paid, we have:

\begin{equation}
\begin{array}{ll}
V_{t_{1}^{+}}&=c_2D(t_1,t_2)\Big[1-\frac{(1-R)\lambda_2+\gamma_2}{\lambda_2+\gamma_2}\Big(1-e^{-(\lambda_2+\gamma_2)(t_2-t_1)}\Big)\Big]\\ 
\end{array}
\end{equation}

Furthermore, we can also define $V_{t_{1}^{-}}$ as the value of the bond at date $t_1^{-}$, immediately before cash flow $c_1$ at $t_1$ is paid, so:

\begin{equation}
\begin{array}{l}
V_{t_{1}^{-}}=c_1+V_{t_{1}^{+}}\equiv A_1\\ 
\end{array}
\label{eq:Bound1}
\end{equation}

Finally, we also consider the value of the collateralized equivalent transaction at date $t_{1}^{-}$, defined as

\begin{equation}
\begin{array}{l}
V_{t_{1}^{-}}^{c}=c_1 + c_2D(t_1,t_2) \equiv B_1\\ 
\end{array}
\label{eq:Bound2}
\end{equation}

When we consider $t\in[t_0,t_1]$ and substitute $V_{t_1}^c$ for $B_1$ in \eqref{eq:FinalPricingEquationBond}, the equation we obtain is very similar to the one that we saw in the single-period example. The same arguments outlined above lead us to the following expression:

\begin{equation}
\begin{array}{l}
D^{-1}(t,t_1)\frac{V_t}{B_1}=\frac{R\lambda_1}{\lambda_1+\gamma_1}+\bar{K}e^{-(\lambda_1+\gamma_1)(t_1-t)}\\
\end{array}
\label{eq:SolvDifEq2}
\end{equation}

We then apply the terminal condition $V_{t_1}=A_1$ to determine $\bar{K}$. Putting all pieces together, we arrive at the final expression:

\begin{equation}
\begin{array}{ll}
V_t=&c_1D(t,t_1)\Big[1-\frac{(1-R)\lambda_1+\gamma_1}{\lambda_1+\gamma_1}\Big(1-e^{-(\lambda_1+\gamma_1)(t_1-t)}\Big)\Big]\\
&+c_2D(t,t_2)\Big[1-\frac{(1-R)\lambda_1+\gamma_1}{\lambda_1+\gamma_1}\Big(1-e^{-(\lambda_1+\gamma_1)(t_1-t)}\Big)\\
&\hspace{20mm}-\frac{(1-R)\lambda_2+\gamma_2}{\lambda_2+\gamma_2}\Big(1-e^{-(\lambda_2+\gamma_2)(t_2-t)}\Big)e^{-(\lambda_1+\gamma_1)(t_1-t)}\Big]\\ 
\end{array}
\label{eq:Bond2FinalPrice}
\end{equation}

By applying sequentially the corresponding terminal conditions, we arrive to the expression for an arbitrary number of periods:

\begin{equation}
\begin{array}{ll}
V_t=&\sum_{p\in I_c}D(t,t_p)c_p\Big[1-\sum_{i=1}^{n}\frac{(1-R)\lambda_i+\gamma_i}{\lambda_i+\gamma_i}\Big(1-e^{-(\lambda_i+\gamma_i)(t_i-t_{i-1})}\Big)\times \prod_{j=1}^{i-1}e^{-(\lambda_j+\gamma_j)(t_j-t_{j-1})}\Big]\\ 
\end{array}
\label{eq:BondGenFinalPrice}
\end{equation}

If we make the intervals defining the piecewise-constant functions tend to zero, $\lambda_i \rightarrow \lambda(s)$ and $\gamma_i \rightarrow  \gamma(s)$, and we have the alternative expression valid for any deterministic functions:

\begin{equation}
\begin{array}{ll}
V_t=&\sum_{p\in I_c}D(t,t_p)c_p\Big[1-\int_{t}^{t_p}((1-R)\lambda(s)+\gamma(s))\exp\Big(-\int_{t}^{s}(\lambda(u)+\gamma(u))du\Big)ds\Big]\\ 
\end{array}
\label{eq:BondGenFinalPriceCont}
\end{equation}

If we define the liquidity-adjusted survival probability 

\begin{equation}
\begin{array}{l}
P^L(t,s) \equiv \exp\Big(-\int_{t}^{s}(\lambda(u)+\gamma(u))du\Big),\\ 
\end{array}
\label{eq:LiqAdjstSurv}
\end{equation}
the equation simply becomes

\begin{equation}
\begin{array}{ll}
V_t=&\sum_{p\in I_c}D(t,t_p)c_p\Big[1-\int_{t}^{t_p}((1-R)\lambda(s)+\gamma(s))P^L(t,s)\,ds\Big]\\ 
\end{array}
\label{eq:BondGenFinalPriceContSimple}
\end{equation}

\subsection{Bond with Absolute Recovery}

\label{app:BondAbsolute}

In the case of a bond with absolute recovery, the payoff upon default is not a recovery fraction of the collateralized equivalent of the bond, but recovery times the notional amount. 

We place ourselves in the same framework as in \ref{app:BondRiskless}. However, besides being deterministic, we now will assume that interest rates $r_j$ also display a piecewise-constant structure, with parameters remaining constant between dates $\{t_j^{r}\}$. Again, we first consider the case in which we only have one period: $[t_0,t_1]$. The situation is the same as in \eqref{eq:OnePeriodBond}, where, using the new recovery convention, we substitute $R \rightarrow R\frac{N}{V_t^c}$. This yields:

\begin{equation}
\begin{array}{ll}
V_t&=D(t,t_1)c_1-\gamma_{1}\int_{s=t}^{t_1}e^{-\lambda_1(s-t)}D(t,s)V_sds\\
&-\int_{s=t}^{t_1}\lambda_1(D(s,t_1)c_1-RN)e^{-\lambda_1(s-t)}D(t,s)ds\\
\end{array}
\end{equation}

Proceeding as in \ref{app:BondRiskless} and imposing the terminal condition $V_{t_1}=c_1$, it needs to be

\begin{equation}
\begin{array}{ll}
V_t&=c_1D(t,t_1)e^{-(\lambda_1+\gamma_1)(t_1-t)}+\frac{\lambda_1RN}{r_1+\lambda_1+\gamma_1}\Big(1-D(t,t_1)e^{-(\lambda_1+\gamma_1)(t_1-t)}\Big)\\ 
\end{array}
\label{eq:Bond1FinalPriceRisky}
\end{equation}
In general, for an arbitrary number of periods, the expression would be

\begin{equation}
\begin{array}{ll}
V_t=&\sum_{p\in I_c}c_pD(t,t_p)\exp\Big(\sum_{i=1}^{p}-(\lambda_i+\gamma_i)(t_i-t_{i-1})\Big)+\\
&RN\sum_{i=1}^{p}\frac{\lambda_i}{r_i+\lambda_i+\gamma_i}\Big(1-D(t_{i-1},t_i)e^{-(\lambda_i+\gamma_i)(t_i-t_{i-1})}\Big)\prod_{k=1}^{i-1}D(t_{k-1},t_k)e^{-(\lambda_k+\gamma_k)(t_k-t_{k-1})}\\ 
\end{array}
\label{eq:BondGenFinalPriceRisky}
\end{equation}

Again, we make the intervals defining the piecewise-constant functions tend to zero, $r_i \rightarrow r(s)$, $\lambda_i \rightarrow \lambda(s)$ and $\gamma_i \rightarrow (s)$, and, making use of the definition \eqref{eq:LiqAdjstSurv}, we have the alternative expression:

\begin{equation}
\begin{array}{l}
V_t=\sum_{p\in I_c}\Big[c_pD(t,t_p)P^L(t,t_p)+RN\int_{t}^{t_p}\lambda(s)D(t,s)P^L(t,s)\,ds\Big] \ .\\ 
\end{array}
\label{eq:BondGenFinalPriceContRiskyLiq}
\end{equation}


\newpage

\begin{thebibliography}{99}

\bibitem{BlackScholes}Black, Fischer and Myron Scholes (1973), "The Pricing of Options and Corporate Liabilities."  \emph{Journal of Political Economy}, Vol. 81, No. 3, May-June, pp. 637-654
\bibitem{Merton}Merton, Robert (1973), "Theory of Rational Option Pricing." \emph{Bell Journal of Economics and Management Science}, 4, Spring, pp. 141-183
\bibitem{HullWhite2013a}Hull, John, and Alan White. "Libor vs. OIS: The derivatives discounting dilemma." \emph{Journal of Investment Management}, Vol. 11, No. 3, 14-27 (2013)
\bibitem{Gregory2009}Gregory, Jon. "Being Two-faced over Counterparty risk." \emph{Risk Magazine}, 22(2) (2009): 86-90
\bibitem{Brigo2009}Brigo, Damiano, Andrea Pallavicini, and Vasileios Papatheodorou. "Bilateral counterparty risk valuation for interest-rate products: impact of volatilities and correlations." {\tt \href{http://arxiv.org/abs/0911.3331}{arXiv:0911.3331}} (2009)
\bibitem{Gregory2012}Gregory, Jon. \emph{Counterparty Credit Risk and Credit Value Adjustment: A Continuing Challenge for Global Financial Markets}, Second Edition. John Wiley \& Sons, 2012
\bibitem{KenyonStamm2012}Kenyon, Chris, and Roland Stamm. \emph{Discounting, Libor, CVA and Funding: Interest Rate and Credit Pricing.} Palgrave Macmillan, 2012
\bibitem{Brigo2013}Brigo, Damiano, Massimo Morini, and Andrea Pallavicini. \emph{Counterparty Credit Risk, Collateral and Funding: With Pricing Cases for All Asset Classes}. John Wiley \& Sons, 2013
\bibitem{ShleiferVishny}Shleifer, Andrei, and Robert W. Vishny. "The limits of arbitrage." \emph{The Journal of Finance} 52.1 (1997): 35-55
\bibitem{Blanco2005}Blanco, Roberto, Simon Brennan, and Ian W. Marsh. "An empirical analysis of the dynamic relation between investment-grade bonds and credit default swaps." \emph{The Journal of Finance} 60.5 (2005): 2255-2281
\bibitem{Jankowitsch}Jankowitsch, Rainer, Rainer Pullirsch, and Tanja Ve\v{z}a. "The delivery option in credit default swaps." Journal of Banking \& Finance 32.7 (2008): 1269-1285
\bibitem{Garleanu}Garleanu, Nicolae, Lasse Heje Pedersen, and Allen M. Poteshman. "Demand-based option pricing." \emph{Review of Financial Studies} 22.10 (2009): 4259-4299
\bibitem{Augustin}Augustin, Patrick. "Squeezed everywhere: Can we learn something new from the CDS-Bond Basis?" Working Paper, 2012
\bibitem{MoriniPrampolini2010}Morini, Massimo, and Andrea Prampolini. "Risky funding: A unified framework for counterparty and liquidity charges." Available at {\tt \url{http://ssrn.com/abstract=1669930}} (2010)
\bibitem{PallavaciniBrigo2011}Pallavicini, Andrea, Daniele Perini, and Damiano Brigo. "Funding Valuation Adjustment: a consistent framework including CVA, DVA, collateral, netting rules and re-hypothecation." {\tt \href{http://arxiv.org/abs/1112.1521}{arXiv:1112.1521}} (2011)
\bibitem{PallavaciniBrigo2012}Pallavicini, Andrea, Daniele Perini, and Damiano Brigo. "Funding, Collateral and Hedging: uncovering the mechanics and the subtleties of funding valuation adjustments." {\tt \href{http://arxiv.org/abs/1210.3811}{arXiv:1210.3811}} (2012)
\bibitem{Piterbarg}Piterbarg, Vladimir. "Funding beyond discounting: collateral agreements and derivatives pricing.", \emph{Risk} 23.2 (2010):97
\bibitem{Lou}Lou, Wujiang. "Coherent CVA and FVA with Liability Side Pricing of Derivatives." Available at {\tt \url{http://ssrn.com/abstract=2449109}} (2014) 
\bibitem{Hull2012a}Hull, John, and Alan White. "The FVA debate." \emph{Risk Magazine} 8 (2012)
\bibitem{Hull2012b}Hull, John, and Alan White. "Collateral and credit issues in derivatives pricing." Available at {\tt \url{http://ssrn.com/abstract=2212953}} (2012)
\bibitem{Hull2013b}Hull, John, and Alan White. "Valuing derivatives: Funding value adjustments and fair value." Available at {\tt \url{ http://ssrn.com/abstract=2245821}} (2013)
\bibitem{Ruiz2013}Ruiz, Ignacio, "FVA Demystified.", Working paper (2013)
\bibitem{BurgardKjaer2011}Burgard, Christoph, and Mats Kjaer. "Partial differential equation representations of derivatives with bilateral counterparty risk and funding costs." \emph{The Journal of Credit Risk} 7.3 (2011): 75-93
\bibitem{BurgardKjaer2012}Burgard, Christoph, and Mats Kjaer. "Generalised CVA with funding and collateral via semi-replication." Available at {\tt \url{http://ssrn.com/abstract=2027195}} (2012)
\bibitem{GunFer2014}Gunnesson, Johan, and Alberto Fern\'andez Mu\~noz de Morales. "Recovering from Derivatives Funding: A consistent approach to DVA, FVA and Hedging" {\tt \href{http://arxiv.org/abs/1403.1086}{arXiv:1403.1086}} (2014)
\bibitem{KPMG2013}KPMG. "FVA - Putting Funding into the Equation". Available at {\tt \url{https://www.kpmg.com/Global/en/IssuesAndInsights/ArticlesPublications/Documents/funding-valuation-adjustment.pdf}} (2013)
\bibitem{Garcia2013}Garc\'ia~Mu\~noz, Luis Manuel, "CVA, FVA (and DVA?) with stochastic spreads. A feasible replication approach under realistic assumptions." Available at {\tt \url{http://mpra.ub.uni-muenchen.de/44568}} (2013)
\bibitem{BrigoMercurio}Brigo, Damiano, and Fabio Mercurio. \emph{Interest rate models-theory and practice: with smile, inflation and credit}. Springer, 2007.
\bibitem{Castagna2012} Castagna, Antonio. "On the Dynamic Replication of the DVA: Do Banks Hedge
their Debit Value Adjustment or their Destroying Value Adjustment?" Available at {\tt \url{http://ssrn.com/abstract=1989403}} (2012)
\bibitem{GregoryGerman}Gregory, Jon, and Ilya German. "Closing out DVA?" Working paper (2012)
\bibitem{Carver}Carver, Laurie. "Quants call for ISDA to clarify close-out values." \emph{Risk}, December (2011)
\bibitem{PykhtinRosen}Pykhtin, Michael, and Dan Rosen. "Pricing counterparty risk at the trade level and CVA allocations." Available at {\tt \url{http://ssrn.com/abstract=1782063}} (2010)
\bibitem{Pykhtin}Pykhtin, Michael. "Modeling credit exposure for collateralized counterparties." Journal of Credit Risk 5.4 (2009): 3-27
\bibitem{GreenKenyon2014} Green, Andrew, and Chris Kenyon. "KVA: Capital Valuation Adjustment." {\tt \href{http://arxiv.org/abs/1405.0515}{arXiv:1405.0515}} (2014)
\bibitem{BrigoBuescu2012}Brigo, Damiano, et al. "Illustrating a problem in the self-financing condition in two 2010-2011 papers on funding, collateral and discounting." {\tt \href{http://arxiv.org/abs/1207.2316}{arXiv:1207.2316}} (2012)
\bibitem{KarShreve}Karatzas, Ioannis, and Steven Shreve. \emph{Brownian motion and stochastic calculus} Springer, Vol. 113, 1991


\end{thebibliography}
\end{document}